\documentclass[journal,twoside]{IEEEtran}
\ifCLASSINFOpdf
\else
\fi
\usepackage{xfrac}
\usepackage{amsmath}
\usepackage{multirow}
\usepackage{color}
\usepackage{array}
\usepackage{amssymb}
\usepackage{pifont}
\newcommand{\xmark}{\ding{55}}%
\usepackage{cite}
\DeclareMathOperator*{\argmin}{argmin}
\usepackage{lipsum}
\usepackage{algpseudocode,algorithm}
\usepackage[subtle]{savetrees}

\newtheorem{rem}{Remark}
\newtheorem{theorem}{Theorem}
\newtheorem{prop}{Proposition}
\newtheorem{corol}{Corollary}

\ifCLASSOPTIONcompsoc
\usepackage[caption=false,font=normalsize,labelfon
t=sf,textfont=sf]{subfig}
\else
\usepackage[caption=false,font=footnotesize]{subfi
g}
\fi

\usepackage{stfloats}
\usepackage{etoolbox}

\makeatletter
\patchcmd{\@begintheorem}{\textit}{\textbf}{}{}
\makeatother

\begin{document}
\bstctlcite{IEEEexample:BSTcontrol}


\title{Bit-Interleaved Multiple Access: Improved Fairness, Reliability, and Latency for Massive IoT Networks}
%
%
%

\author{Ferdi~Kara,~\IEEEmembership{Senior~Member,~IEEE,}~Hakan~Kaya,~Halim~Yanikomeroglu,~\IEEEmembership{Fellow,~IEEE,} ~Benjamin~K.~Ng,~\IEEEmembership{Senior~Member,~IEEE,}~and ~Chan-Tong~Lam,~\IEEEmembership{Senior~Member,~IEEE.}
\thanks{The work of F. Kara is financially supported by the Scientific and Technological Research Institution of Türkiye (TUBITAK).}
\thanks{F. Kara is 
with the Computer Engineering, Zonguldak Bulent Ecevit University, Zonguldak, Türkiye. He is also with the Department of Systems and Computer Engineering, Carleton University, Ottawa, K1S 5B6, ON, Canada, email: f.kara@beun.edu.tr.}
 \thanks{H. Kaya is 
with the Electrical-Electronics Engineering, Zonguldak Bulent Ecevit University, Zonguldak, Türkiye,  e-mail: hakan.kaya@beun.edu.tr.}
\thanks{H. Yanikomeroglu is with the Department of Systems and Computer Engineering, Carleton University, Ottawa, K1S 5B6, ON, Canada, e-mail:halim@sce.carleton.ca.}
\thanks{ B. Ng and C-T. Lam are with the Faculty of Applied Sciences, Macao Polytechnic University, Macao SAR, China, e-mail:\{bng, ctlam\}@mpu.edu.mo}
}

\maketitle
\begin{abstract}
Internet-of-Things (IoT) networks require massive connections in dense areas. Therefore, a resource efficient multiple access scheme seems inevitable to enable immense connectivity where multiple devices have to share the same resource block. Non-orthogonal multiple access (NOMA) has been considered as the strongest candidate in recent years. However, in this paper, by considering the practical implementation, we first provide a true power allocation (PA) constraint with finite alphabet inputs for conventional downlink NOMA and demonstrate that it cannot support massive connections in practical systems. To this end, we propose bit-interleaved multiple access (BIMA) scheme in downlink IoT networks. The proposed BIMA scheme implements bitwise multiaccess interleaving and deinterleaving at the transceiver ends and  there are no strict PA constraints, unlike conventional NOMA, thus allowing a high number of devices in the same resource block. We provide a comprehensive analytical framework for BIMA by investigating all key performance indicators (KPIs) to present both information-theoretic  (i.e., ergodic capacity [EC] and outage probability [OP]) and finite alphabet inputs (i.e., bit error rate [BER]) performance metrics with both instantaneous and statistical channel ordering. In addition, we define Jain's fairness index and proportional fairness index in terms of all KPIs. Based on the extensive computer simulations, we reveal that BIMA outperforms conventional NOMA significantly, with a performance gain of up to 20-30 dB in terms of KPIs in some scenarios. In other words, compared to conventional NOMA schemes, the same KPIs are met in BIMA with 20-30 dB less transmit power, which is quite promising for energy-limited use cases. Moreover, this performance gain becomes greater when more IoT devices are supported. BIMA provides a full diversity order for all IoT devices and enables the implementation of an arbitrary number of devices and modulation orders, which is crucial for IoT networks where a huge number of devices should be supported in a single resource block in dense areas. In addition to the overall performance gain, BIMA guarantees a fairness system where none of the devices gets a severely degraded performance and the sum-rate is shared in a fair manner among devices. It guarantees QoS satisfaction for all devices. Finally, we provide an intense complexity and latency analysis for BIMA and demonstrate that it provides lower latency compared to conventional NOMA receivers, since it allows parallel computation at the receivers and no iterative operations are required. We show that compared to conventional NOMA receivers, BIMA reduces latency by up to 350\% for specific IoT devices and 170\% on average.
\end{abstract}

\IEEEpeerreviewmaketitle
\section{Introduction}

Wireless communication technologies have improved considerably in recent decades. The evolution of wireless communications has been driven by the demands of users, customers, and clients and by technological developments. For instance, with the introduction of smart living areas (e.g., cities, agriculture, factories, etc.), wireless access has moved beyond personal communication: it is now called Internet of everything (IoE) \cite{Giordani2020,Tataria2021,Huawei}. In order to enable smart living areas, wireless infrastructure needs to support ultra-dense networks where a massive number of nodes (e.g., sensors, smart-watches, tablets, etc.) have wireless access, which is called massive machine type communication (mMTC)—one of the three major concepts of 5G and beyond. This is driven by the Internet of Things (IoT) \cite{IoT_survey,IoT_survey2}. Recently, the number of IoT devices has increased exponentially. The most recent vendor reports \cite{Ericsson2022} show that 14.6 billion IoT connections exist in wireless infrastructure by 2022 and with a 13\% annual grow rate, it will reach 30.2 billion by 2027. This will be almost equal to 40\% of total wireless access at that time. Furthermore, these IoT networks have small operating areas (e.g., a building and a factory), unlike traditional cellular networks, and these small areas trigger a challenging resource allocation and interference management problem since massive numbers of devices should be served within these small areas. In the wireless  spectrum, physical radio resource blocks (RB) (i.e., time and frequency) are limited and costly. For instance, in 5G new radio (NR) standards \cite{3gpp17}, with a 20 MHz bandwidth (i.e., maximum bandwidth in LTE legacy ), a maximum number of 92 RBs\footnote{Although it is available 106 PRBs, only 92 of them can be assigned to users due to usage of broadcast and control signaling \cite{3gpp17}.} can be allocated to users within a subframe. However, in massive IoT applications (e.g., smart agriculture), thousands of sensor or control nodes may require wireless access. In these cases, whether we need more bandwidth, that is costly or more complex resource allocation (RB scheduling algorithms in layer 2) solutions are required. Therefore, it is not possible to allocate each IoT device to an orthogonal resource to avoid interference, since a massive number of devices need to be served in small, dense areas. For this reason, more than one IoT devices should share a resource block to enable mMTC. In this regard,  non-orthogonal multiple access (NOMA) is seen as a strong candidate for IoT networks \cite{Shirvanimoghaddam2017} since it allows multiple devices to share the same resource blocks by splitting them into the power domain. In this way, the spectral efficiency of the network increases and it becomes possible to serve multiple devices with the number of more than the available resource blocks \cite{Vaezi2018}. Accordingly, NOMA is seen as an enabler for ultra-dense networks, and tremendous efforts have been devoted to integrating NOMA in IoT applications\cite{NOMA_IOT1,NOMA_IOT2,NOMA_IOT3,NOMA_IOT4,NOMA_IOT5,9711564}.


\subsection{Related Work and Motivation}
In the literature, NOMA schemes have been shown to be spectrally efficient and superior to orthogonal multiple access schemes in terms of capacity and outage performance \cite{Saito2013,Ding2014} and their implementation with recent technologies (e.g., integrated terrestrial-satellite communications \cite{9729897} ) has attracted a remarkable attention. However, in these capacity and outage performance evaluations, a perfect successive interference canceler (SIC) is generally assumed \cite{Saito2013,Ding2014,9729897,Sun2015,Liu2015b,Ding2016} or an imperfect SIC effect is modeled as a coefficient (constant or Gaussian random variable) \cite{Zhong2016,Kader2017,Im2019,Kara2020}. Moreover, capacity and outage performances have been analyzed only in terms of information-theoretic perspectives where the transmitted signals are assumed to have infinite alphabets (a.k.a., Gaussian inputs); therefore, the transceivers/baseband blocks have not been considered in those analysis. However, in practical scenarios, baseband signal processing techniques (e.g., an in-phase and quadrature [IQ] modulation and demodulation) are required for wireless communications. Therefore, the practical systems take their values from finite alphabet inputs such as a M-level quadrature-amplitude modulation (M-QAM) constellation. In this regard, once a baseband transceiver is implemented, it has been shown that the imperfect SIC cannot be modeled only by a coefficient, and NOMA networks have a degradation of error performance (bit error rate, BER) for both downlink \cite{9552864,Kara2018f} and uplink \cite{Kara2018f,Wang2017} scenarios. These studies led other researchers to investigate the BER performance of NOMA systems with finite alphabet inputs to reflect their practical implementations. Then, it is not as tremendous as ergodic rate and outage probability analysis, but a considerable amount of studies are devoted to investigate the BER performance of NOMA networks for various fading channels in both downlink and uplink scenarios \cite{Bariah2018,Assaf2019,Lee2019,Kara2019,Kara2019a,Garnier2020,NOMA_BER, 9252989, 8788603,9509752,9763842}. In particular, the authors in \cite{Kara2018f} showed that the SIC detector suffers from the error floor in uplink NOMA and may not be the optimal detection algorithm for uplink NOMA unlike downlink NOMA schemes. Then, the joint maximum-likelihood detector is proposed for uplink NOMA and it is proved to be optimal solution and is capable of error floor in uplink NOMA \cite{8788603,9509752,9763842}. On the other hand, in the BER analysis of the downlink NOMA studies \cite{Kara2018f,Bariah2018,Assaf2019,Lee2019,Kara2019,Kara2019a,Garnier2020,NOMA_BER, 9252989}, the number of users\footnote{User refers to an IoT device in this paper. Thus, the terms ``user'' and ``(IoT) device'' are used interchangeably throughout this paper.} is limited to only two or three. This is because the BER performance of users degrades (e.g., some users may have BER value of 1) as the number of users increases due to SIC operations, although NOMA can serve more users theoretically. Moreover, in most studies \cite{Kara2018f,Bariah2018,Assaf2019,Lee2019,Kara2019,Kara2019a,Garnier2020,NOMA_BER}, the modulation orders in the transceiver have been selected as binary phase-shift keying (BPSK) and/or quadrature phase-shift keying (QPSK). If the modulation orders increase in the NOMA schemes, the BER performance of the devices become worse. Furthermore, the higher the modulation orders, the more difficult it becomes to detect symbols at the receivers; this is because the total constellation after superposition at the transmitter scatters too much. Indeed, this scattering causes a non-detectable constellation and a conceptual flaw in downlink NOMA schemes. Therefore, the power allocation (PA) constraint becomes too strict in practical scenarios and, contrary to PA optimization studies in terms of various constraints \cite{Liu2016e,Lei2016,Zhu2017,Oviedo2019}, no device can detect symbols if this strict PA constraint is not satisfied. In \cite{Lee2019}, it was shown that the PA of one of the users should be higher than $0.9$ even in a case involving two users with $16$-QAM. The strict PA constraint triggers another problem, namely user unfairness, where the performance of nodes is affected dramatically and one user experiences degraded performance. To address this, some studies \cite{Fairness1,Fairness2,Fairness3,Di2016,Gui2019,Liu2020} have examined fairness in NOMA networks, and a few algorithms have been proposed to this end. However, as in almost all NOMA studies, only information-theoretic issues were considered, and no practical implementation issues (e.g., constellation-based constraints for finite alphabet inputs) were discussed. In addition, the aforementioned studies \cite{Fairness1,Fairness2,Fairness3,Di2016,Gui2019,Liu2020} are also limited by only considering two-user networks. This strict PA constraint (i.e., unfairness) also manifests itself with increasing numbers of users; therefore, in three-user \cite{Bariah2018,Assaf2019} and four-user\cite{Aldababsa2020} NOMA networks, the authors consider QPSK and BPSK, respectively. Neither the modulation order can be increased in a three-user network nor the number of served users in a single resource block can be improved with any modulation order.
 The aforementioned fairness evaluations trigger another discussion that is the reasonable signal-to-noise ratio (SNR) regime to implement NOMA. Due to intentionally created inter-user-interference (IUI), the performance of NOMA systems are SNR-limited. In other words, the NOMA schemes can outperform OMA counterparts (i.e., interference-free) in only very low SNR regions.  In NOMA schemes, one (or all) users have an interference-limited performance (i.e., error floor or capacity upper bound) where its (their) performance is not improved even though the received SNR is increased. Therefore, NOMA is not suitable for indoor IoT applications where the SNR is generally high due to low distances from AP to IoT devices.

As explained above, NOMA networks lack high reliability, since all users experience a degradation in BER performance due to IUI. For this reason, many studies have been devoted to improving the BER performance of NOMA for uplink \cite{Ye2018a,Lin2019b,Ng2018} and downlink \cite{Ng2018,Chang2018,Wan2020,Ozyurt2020} networks. These studies are mostly based on constellation designs with rotation and/or phase shift. In \cite{Ye2018a}, the authors proposed to rotating the constellation in a two-user uplink NOMA scheme. The achievable rate analysis was provided for the proposed system, and the PA was optimized using the variational approximation method. The BER simulations were also presented for QPSK in the scenarios considered, and a subtle performance gain was observed. Then, in \cite{Lin2019b}, the authors investigated the optimum inter-constellation rotation based on minimum Euclidean distance for a two-user uplink NOMA. They provided an SNR-independent optimization in addition to eliminating the error floor of error performance. Another study focused on improving BER performance was undertaken by the authors in \cite{Ng2018}, who proposed joint power and rotation optimization for MIMO-NOMA with two users and 4-QAM for both uplink and downlink cases. Their approach demonstrated better BER performance than traditional NOMA schemes. NOMA with phase rotation was proposed in \cite{Chang2018} where joint multi-device detection was also proposed for two-user downlink NOMA. The proposed scheme was evaluated with convolutional codes and SIC-based soft decision Viterbi decoding. The authors presented BER performance improvements for the QPSK-QPSK and QPSK-16-QAM constellations. Unlike constellation rotation, an interference alignment and independent component analysis-based semi-blind two-user downlink NOMA scheme was proposed in \cite{Wan2020} where the phase of one of the devices was arranged according to the other user's symbols. The proposed scheme was evaluated for 4-QAM (i.e., QPSK) constellations, and the symbol error rate (SER) performance was demonstrated by simulations. However, all aforementioned studies only considered two-user networks and lower modulation orders for the devices. To the best of the authors' knowledge, the only study with more than two users (i.e., four and six) was the work in \cite{Ozyurt2020} where the authors apply a coordinate (a.k.a., component) interleaving for the
$\sfrac{\pi}{4}$ rotated 4-pulse amplitude modulation (4-PAM) signals which is inherited from the signal space diversity technique in literature \cite{681321}. In \cite{Ozyurt2020}, 4-PAM signals of each users are rotated with $\sfrac{\pi}{4}$ on counter clockwise and then in-phase and quadrature components of each IQ symbols are interleaved among I-Q coordinates  (e.g., in a two user scenario, both components of the first user's symbol are mapped into only in-phase whereas both components of the second user's symbol are mapped into only quadrature domain with related PA coefficients). Therefore, half of the devices were assigned to an in-phase domain, while the other half was assigned to a quadrature domain, so that the IUI decreased by half in each domain. However, the modulation orders of the devices are still limited by 4-PAM, and the proposed scheme in \cite{Ozyurt2020} cannot be used with two-dimensional modulation schemes, so it is unlikely to be implemented in standards.

On the other hand, the performance improvement studies in NOMA \cite{Ye2018a,Lin2019b,Ng2018,Chang2018,Wan2020,Ozyurt2020} aim to optimize the constellation after superposition coding. In other words, they try to obtain an optimized super-constellation for multi-user cases. This reminds us another efficient technique used in digital video broadcasting (DVB) communication called hierarchical modulation (HM) \cite{Weck2003}, a.k.a, embedded modulation\cite{Kannan1993}. In HM, the data sequence is divided into two parts and named as high priority (HP) and low priority (LP) bits. Then, according prioritizing, a merged super-constellation is generated at the transmitter. In DVB standards \cite{ETSI2015}, the HP has 2 bits and LP has 4 bit; therefore, in HM, a 64-QAM constellation is generated where 2 HP bits define the each quadrant in I-Q domain wheres 4 LP bits map the exact point in each quadrant. In this regard, the HP bits can be recovered with QPSK demodulator whereas a 64-QAM demodulator is needed for LP bits. The performance of HM has been well-investigated by researchers \cite{Vitthaladevuni2003,Hossain2006,Barmada2005,Meric2013}. However, its usage is limited in DVB and not considered in other applications. Besides, there are also some limitation for HM. The HM is mainly considered as a data partition according to their importance (e.g., entropy) rather than as a multiple access scheme. The usage as a multiple access scheme is limited by only 2 DVB subscribes. Besides, there is a severe unfairness in HM since the HM is created according to prioritizing level. E.g., LP data may have up to 15 dB performance loss compared to HP data. This unfairness issue limited the usage of HM in DVB rather than mission critical wireless applications where the reliability is crucial; thus, preventing researchers from HM in wireless applications where the channel impairments already causes performance loses. 

In addition to user unfairness and BER performance degradation, the other disadvantage of NOMA is the latency at the receivers due to iterative SIC decoding. Hence, a NOMA design without SIC detection was proposed in \cite{Qiu2019a} and \cite{Qiu2019} for slow fading and block fading channels, respectively. In \cite{Qiu2019a}, a lattice-partition-based model was proposed, and for a two-user downlink NOMA, the proposed scheme was appropriate for a wide range of  outage probability targets regions. Along similar lines, in \cite{Qiu2019}, the authors used an algebraic rotating lattice design, which allowed the receivers to be reduced to single-device detection without an SER performance degradation. In both schemes \cite{Qiu2019a,Qiu2019}, performance evaluations were presented for only two-user networks and the given methods can not be extended for larger networks. Although iterative operations have not been completely eliminated in \cite{Ozyurt2020}, the required SIC operations are halved as interference in each domain is limited by half. However, in multi-device networks, the receiver latency still remains to be resolved.

As we can see from the above discussions, NOMA suffers from BER degradation caused by IUI. Due to the PA constraints, users encounter dramatic unfairness, where one user may experience severe performance limitations, although overall performance seems to be improved. Moreover, due to iterative SIC operations, the receivers have high latency. Evidently, conventional NOMA networks cannot be an optimal solution for IoT networks due to BER degradation, user unfairness, and high receiver latency. Therefore, an alternative multiple access scheme is clearly needed to enable IoT networks for when a massive number of devices need service in an ultra-dense area, such that multiple devices are allowed in a single resource block. To this end, in this paper, we propose bit-interleaved multiple access (BIMA) to enable massive connection in IoT networks. The proposed BIMA overcomes all problems in NOMA, e.g., latency, unfairness, and reliability, etc., where all devices have better and fair performances compared to conventional NOMA. Furthermore, with the proposed BIMA, none of the devices requires a SIC detector, so it has lower receiver latency.

\subsection{Contributions}
The main contributions of this paper are summarized as follows.
\begin{itemize}
\item Firstly, for practical implementations of conventional NOMA, we present a true PA constraint with finite alphabet inputs of M-QAM constellations to ensure reliable signal detection at the receivers. In doing so, we prove that conventional NOMA can not support multiple devices (not more than three or four) in a single resource block. Therefore, it cannot be a true candidate for ultra-dense networks such as IoT applications.
\item For downlink of the dense IoT networks, we propose BIMA, where the information of the devices are concatenated in bitwise by using multiaccess interleaver (MI) rather than symbol-wise superposition coding like conventional NOMA. Thanks to MI at the transmitter and deinterleaving at the receiver ends, the overall multiaccess communication is completed by a single IQ signal. In BIMA, we create an additional orthogonal dimension (i.e., constellation points) to distinguish users' signals (likewise modulation portioning or hierarchical modulation). Therefore, it can be considered as the fourth (constellation) dimension to allow orthogonal multiple access (along with time, frequency, and code division multiple access schemes). Nevertheless, the BIMA distinguishes from OMA legacy since it requires only one physical radio resource. Therefore, it is a single resource multiple access scheme likewise NOMA. In BIMA, since an interference-free transmission is achieved and no interference mitigation technique (i.e., SIC) is required at the receivers. We also demonstrate that there is no power constraint with the proposed BIMA.

\item We analyze the proposed BIMA in terms of all key performance indicators (KPIs). We derive closed-form expressions of ergodic capacity (EC), outage probability (OP), and bit error rate (BER) for both instantaneous and statistical channel ordering schemes. All derived expressions are validated via computer simulations. 
\item We also discuss fairness in BIMA and demonstrate that BIMA offers higher fairness in terms of all KPIs compared to conventional NOMA schemes. To assess the fairness, we first define the well-known Jain's fairness index and then, we propose two modified Jain's fairness indexes to evaluate reliable communication rather than only sum rate. We also introduce a proportional fairness index in terms of all KPIs to demonstrate the performance difference among devices.
\item Based on extensive simulations, we show that BIMA outperforms conventional NOMA in terms of all KPIs. For all KPIs, the BIMA offers a monotonously increasing performance w.r.t. transmit power that is not the case in conventional NOMA where error floors or upper bounds occurs due to IUI. In some cases, BIMA saves up to 20-30 dB transmit power to achieve the same KPIs with conventional NOMA which is a significant gain in especially energy-limited use-cases. Besides, in both instantaneous and statistical channel ordering cases, we show that BIMA offers the full diversity order for all users in terms of all KPIs. Furthermore, BIMA has no limiting constraint for either the number of devices or the modulation order. In the numerical evaluation, we also show that BIMA always offers fair communication and converges to the maximum value in a high SNR regime.
\item We also analyze the receiver complexity and latency of BIMA and present comparisons with conventional NOMA (with SIC receivers). We demonstrate that BIMA provides much less receiver latency. Besides, this gain increases with the number of the devices. 
\end{itemize}
\subsection{Notation}
Throughout this paper,  $\left|.\right|$ denotes the absolute value of a scalar/vector and $\binom{.}{.}$ is used for the binomial coefficient. $\hat{\ }$ denotes the estimated symbol or index. $\mathsf{CN}(\mu,\sigma)$ is a complex Gaussian distribution that has independent real and imaginary random variables with the $\mu$ mean and the $\frac{\sigma}{2}$ variance. $\mathrm{Q}(z)$ is the Marcum-Q function, which is given by $\mathrm{Q}(z)=\int\limits_z^\infty\frac{1}{\sqrt{2\pi}}\exp(-\sfrac{z^2}{2})dz$.
\subsection{Organization}
The rest of the paper is organized as follows. In Section II, we present system models for both conventional NOMA and the proposed BIMA. The detection at the receivers, the signal-to-interference plus noise ratio (SINR) definitions, and the channel ordering schemes are also given in this section. Then in Section III, theoretical analysis for all KPIs is presented and the closed-form expressions are derived. In Section IV, we define Jain's fairness index and proportional fairness index in terms of all KPIs to present a comprehensive evaluation.  In Section V, computer simulations are presented to validate the theoretical analysis and to compare with benchmarks. In Section VI, receiver complexity and latency analyses are conducted, and comparisons with conventional NOMA are presented. Finally, Section VII concludes the paper with discussions and considerations for future work.
\begin{figure*}
		\centering
    \includegraphics[width=.85\textwidth]{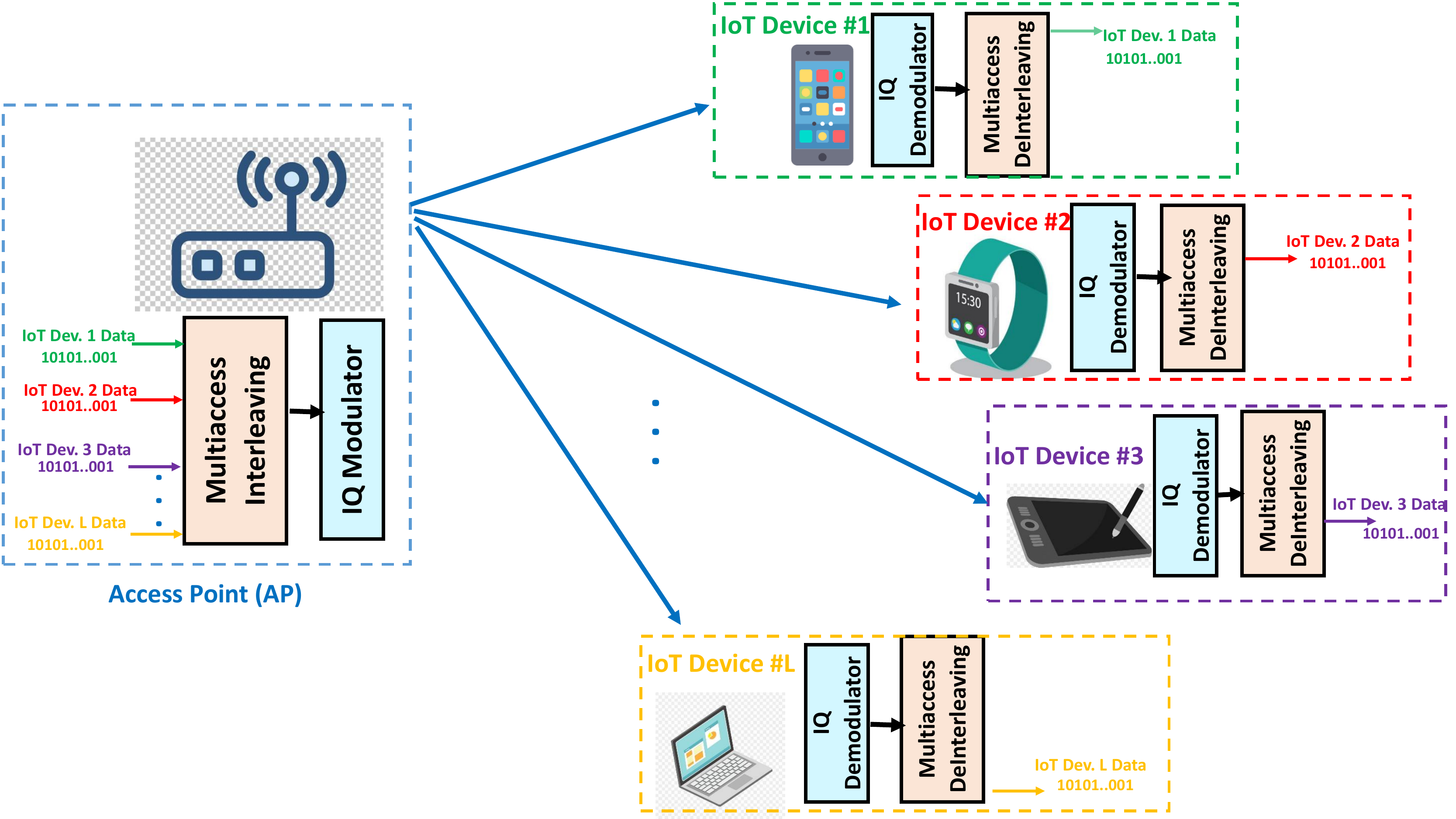}
    \caption{The illustration of  an IoT network enabled by BIMA.}
    \label{sys_model}
\end{figure*}
\section{System and Channel Models}
In this paper, we consider a downlink scenario where a transmitter--access point (AP)-- sends data to $L$ IoT devices, as shown in Fig. 1. All nodes are assumed to have a single antenna\footnote{To make it easier to follow for readers, in this paper, we present BIMA with a single-input single-output (SISO) case. However, BIMA could be implemented in any antenna configuration, e.g., single-input-multiple-output (SIMO), multiple-input single-output (MISO), multiple-input multiple-output (MIMO) or it can be easily used with other physical layer techniques. The interplay between BWNOMA and other physical layer techniques (e.g., MIMO or cooperative communication) is considered as a future work. Besides, the home devices in the Fig. 1 are given just for better illustration. However, the proposed BIMA is not application-limited and, any kind of smart devices (e.g., smart factory devices, human body network devices, etc.) can be supported by BIMA.}, and the channel fading coefficient between each node is a flat-fading Rayleigh channel. To enable the massive connectivity within a resource-limited dense network, a resource-efficient multiple access (e.g., (NOMA) scheme) is required; thus, the AP transmits all devices' symbols at the same resource block (i.e., frequency, time, code). 

In order to present a comprehensive performance evaluation and comparison, we define both conventional NOMA and the proposed schemes in the following. We chose NOMA as the benchmark scheme since it is seen as the strongest candidate for IoT networks to support massive connectivity since the number of devices is much more than available orthogonal radio resource blocks. 

\subsection{Conventional NOMA}
In the conventional NOMA scheme, the AP implements a superposition coding after modulating all devices' symbols. To this end, the total superposition coded symbol at the transmitter becomes
\begin{equation}
    x_{sc}=\sum_{i=1}^L\sqrt{\alpha_i}x_i,
\end{equation}
where $\alpha_i$ is the power allocation (PA) coefficient for the $i$th IoT device. Without loss of the generality, we assume $\alpha_i<\alpha_{i+1}$ and $\sum_{i=1}^L \alpha_i=1$. In (1), $x_i$ is the baseband modulated symbol of $i$th IoT device with $M_i$-ary modulation. $E\left[|x_i|^2\right]=1, \ \forall i$. The received signal at the each IoT device is given by
\begin{equation}
    y_i=\sqrt{P}x_{sc}h_i+n_i, \ i=1,2,\dots,L,
\end{equation}
where $P$ is the transmit power of the AP. $h_i$ and $n_i$ are the flat fading channel coefficients between the AP and $i$th IoT device and the additive Gaussian noise at the receiver $i$, respectively. $h_i\sim \mathsf{CN}(0,\sigma_i^2)$ and $n_i\sim \mathsf{CN}(0,N_0)$ are defined.
\subsubsection{Detection at the IoT devices}
Since the PA coefficient of the $L$th IoT device is greater than any of the others, the $L$th IoT device detects its own symbols by pretending the symbol of the other IoT devices as noise. Thus, the maximum likelihood (ML) detection at the $L$th IoT device is given as
\begin{equation}
    \hat{x}_L=\argmin_{k}{\left|y_L-\sqrt{P\alpha_L}h_Lx_{L,k}\right|^2}, \ k=1,2,\dots, M_L,
\end{equation}
where $x_{L,k}$ denotes the $k$th point in the $M_L$-ary constellation.

However, the $i$th IoT device needs to implement an iterative successive interference canceler (SIC)\footnote{Please note that in conventional NOMA schemes, a joint maximum-likelihood detection (JMLD) can be also implemented where an exhaustive search is performed to obtain all users' symbols at once rather than iterative SIC algorithm. However, in the open literature, it is proven that in the downlink NOMA, SIC and JMLD have exactly the same performance \cite{9252989} unlike in the uplink JMLD outperforms SIC \cite{8788603,9509752,9763842}. Therefore, in this paper, we also perform SIC for conventional NOMA since the majority of downlink NOMA schemes consider SIC. } to detect its own symbols such that it first detects all $j$th IoT devices' symbols (i.e., $j=i+1,i+2,\dots,L$) and subtracts these symbols from the received signal. Therefore, the detection process at the $i$th IoT device is given by
\begin{equation}
    \hat{x}_i=\argmin_{k}{\left|y_i^{(L-i+1)}-\sqrt{P\alpha_i}h_ix_{i,k}\right|^2}, \ k=1,2,\dots, M_i,
\end{equation}
where
\begin{equation}
    y_i^{(L-i+1)}=y_i^{(L-j+1)}-\sqrt{P}h_i\sqrt{\alpha_j}\hat{x}_j, \ j=i+1,i+2,\dots,L,
\end{equation}
where $y_i^{(1)}\triangleq y_i$ and
\begin{equation}
    \hat{x}_j=\argmin_{k}{\left|y_i^{(L-j+1)}-\sqrt{P\alpha_j}h_ix_{j,k}\right|^2}, \ k=1,2,\dots, M_j.
\end{equation}
\subsubsection{Power Allocation Constraint}
In the literature, great efforts have been devoted to optimizing the PA coefficient under different constraints, such as sum-rate maximization, outage probability, energy efficiency, etc., \cite{Liu2016e,Lei2016,Zhu2017,Oviedo2019}. To the best of the authors' knowledge, all of the aforementioned studies optimize PA by considering the theoretical Shannon limit, and no transceiver (i.e., baseband signal processing) is considered. However, in practical systems, the transmitted signals take their values from finite alphabet inputs such as M-QAM constellation alphabet. These aforementioned PA optimization algorithms do not consider practical systems (e.g., finite alphabet inputs). For instance, with an M-QAM alphabet, they do not guarantee successful decoding at the receivers. While they may appear optimal from an information-theoretic perspective, the IoT devices can not detect their symbols when transceivers are implemented. Therefore, for practical applications, a PA constraint should be defined to ensure successful decoding for IoT devices. To this end, we revisit \cite{kara2020mttw}, and according to the ML decision rule for detecting symbols correctly with finite alphabet inputs, the practical PA constraint is given as follows.
\begin{theorem}
Regardless of the channel ordering, in conventional NOMA, to make it practical (i.e., with finite alphabet inputs), the PA coefficient of the $i$th user should be higher than weighted sum of the $j$th user's PA coefficients $\forall j$, $j=1, 2,\dots,i-1$. Otherwise, with an M-ary modulation-demodulation implementation, none of the signals becomes detectable at the receivers. This detectable PA constraint includes M-ary modulation order value for each user and is given as
\begin{subequations} 
\begin{eqnarray}
   & \alpha_i>(M_i-1)\left(\sum_{j=1}^{i-1}\sqrt{\frac{\alpha_j}{M_j-1}}(\sqrt{M_j}-1)\right)^2, \\
   &\text{s.t} \ \alpha_1>0, \\
    &\text{s.t} \ \sum_{i=1}^L\alpha_i=1,
\end{eqnarray}  
\end{subequations}
where (7a) ensures a detectable signal design at the receivers by SIC detectors given in (4)-(6), (7b) ensures a positive PA value for each user and (7c) limits the total transmit power.
\end{theorem}
\begin{IEEEproof}
Please see Appendix A.
\end{IEEEproof}
\subsubsection{Signal-to-Interference-plus-Noise Ratio (SINR)}
Since the symbols of the IoT devices are conveyed through a broadcast channel (simultaneously on the same resource block), the IoT devices encounter an inter-user-interference (IUI). When considering the decoding processes, the signal-to-interference plus noise ratio for the $i$th IoT device is given by
\begin{equation}
  \mathsf{SINR}_i^{(\mathsf{conv})}= \frac{\rho\alpha_i|h_i|^2}{\rho|h_i|^2\sum\limits_{j=i+1}^L\alpha_j\delta_j+\rho|h_i|^2\sum\limits_{p=1}^{i-1}\alpha_p+1}, 
\end{equation}
where  $\rho=\sfrac{P}{N_0}$ and $\delta_j\triangleq|x_j-\hat{x}_j|^2$ are defined. In (8), the first and second terms in the denominator denote the effects of the imperfect SIC and IUI of the devices with lower SIC orders, respectively. We should note that the imperfect SIC effect used in this paper is a more accurate model than those used in existing studies, since it includes the actual difference between the transmitted and detected symbols of other IoT devices rather than defining a constant and/or independent random variable.  

\subsection{ Bit-Interleaved Multiple Access (BIMA)}
Although conventional NOMA is capable of serving multiple IoT devices, the PA constraint for successful decoding (7) limits the number of devices. Otherwise, none of the IoT devices can detect their own symbols since there is no PA coefficient that satisfies (7) with multiple IoT devices. Also, even in the case of two IoT devices, the constellation of IoT devices is limited by $16$-QAM; otherwise, the IoT devices cannot detect symbols with higher modulation orders. To overcome this constraint, a new resource-efficient multiple access technique is required to enable massive connection in dense areas. In this regard, we propose bit-interleaved multiple access (BIMA) where we implement a bitwise multiaccess interleaving and deinterleaving at the transceiver ends. In the BIMA, the information bits of all IoT devices combined in bitwise (i.e., concatenated and interleaved) as a new data sequence. It is given as
\begin{equation} 
    b_{bw}=\mathrm{interlv}\left([b_1,b_2,\dots,b_L]\right),
\end{equation}
where $\mathrm{interlv} ()$ denotes random interleaving. $b_i$ is the information bits of the $i$th IoT device and each of them has $\log_2(M_i)$ bits. Therefore, the total interleaved data has $\sum\limits_{i=1}^L\log_2M_i$ bits. The data is then modulated\footnote{This paper proposes BIMA as a new multiple access technique for IoT networks. Therefore, to focus on the conceptual design in deeply, we do not present error-correcting channel coding here. However, as being in all practical systems, a channel coding can be implemented before modulation which will definitely improve the BER performance of BIMA. Nevertheless, please note that coding gain has a similar effect in all communications systems since it is related to bit correction according to used channel coding algorithm and it does not depend on the multiple access schemes. The performance improvement with the coded systems is clear to a researcher in the field; thus, is beyond the scope of this paper.} by an IQ modulator and transmitted to the IoT devices simultaneously. Since all users' information is transmitted in a single IQ symbol without any interference, the proposed BIMA can be seen as an orthogonal multiple access scheme. However, we create a fourth dimension (i.e., interleaved constellation points) in addition to physical orthogonal radio resources (i.e., time and frequency). To this end, the proposed BIMA is an orthogonal multiple access scheme. However, it differs from OMA legacy since it uses only one radio resource whereas the other OMA schemes require multiple physical radio resources for multiple access. In this regard, BIMA can have an important role when the physical radio resources are not enough. For instance, in ultra-dense networks (e.g., IoT applications), the BIMA has a great potential since the number of devices vastly exceeds the number of available radio resources.

The signal received at each IoT device is given by
\begin{equation}
    y_i=\sqrt{P}x_{bw}h_i+n_i, \ i=1,2,\dots,L,
\end{equation}
where $x_{bw}$ is the modulated symbol by $M_{bw}$-ary modulation. $b_{bw}$ is mapped into $x_{bw}$ which is a point in a $M_{bw}$-ary modulation constellation. $M_{bw}=\prod\limits_{i=1}^LM_i$ and $E\left[|x_{bw}|^2\right]=1$. 
The obtained $M_{bw}$-ary constellation in BIMA has the same size as the total constellation that is obtained after superposition coding in NOMA (1). However, the total constellation in NOMA has an irregular structure due to power allocation that may cause an undetectable signal as already discussed in Theorem 1 and Appendix A. On the other hand, BIMA has a regular $M_{bw}$-ary constellation so that a simple detection algorithm can be used as being in all interference-free (e.g., OMA) schemes.
\subsubsection{Detection at the IoT receivers}
Since the information bits of the IoT devices are combined bitwise and transmitted as a single IQ symbol, any of the IoT devices can detect the combined symbol with a simple ML detector. Therefore, no iterative SIC decision is required at any device. In the ML detector, the received signal at each IoT device is compared with the $M_{bw}$-ary modulation constellation, and the point that has the minimum Euclidean distance is decided as the detected symbol. The ML detection for BIMA is given as
\begin{equation}
\begin{split}
  \hat{x}_{bw}=\argmin_{k}{\left|y_i-\sqrt{P}h_ix_{bw,k}\right|^2},&\\ \ k=1,2,\dots, M_{bw}, \ i=1,2,\dots,L,& \end{split}
\end{equation}
where $x_{bw,k}$ denotes the $k$th point in the $M_{bw}$-ary constellation. Then, the estimated IQ symbol (i.e., $\hat{x}_{bw}$) is de-mapped to $\hat{b}_{bw}$ according to $M_{bw}$-ary modulation constellation where $\hat{b}_{bw}$ is the bitwise representation of the detected symbol.

Next, each IoT device implements a deinterleaving operation to obtain its own information bits. This is given as
\begin{equation}
    b_{i}=\mathrm{deinterlv}\left(\hat{b}_{bw}\right),
\end{equation}
where $\mathrm{deinterlv}()$ is the multiaccess deinterleaving operation at each IoT device. It is assumed that IoT devices have knowledge of the interleaving array used at the transmitter. After the deinterleaving process, each IoT device extracts their own binary bits from the whole data sequence. Thus, it is called multiaccess interleaving (MI). This MI guarantees fairness in the proposed model; otherwise, without an interleaving if we only concatenate data sequentially, the information bits of one or more IoT devices may always be mapped into adjutant points in the $M_{bw}$-ary constellation, and those IoT device(s) may have poor performance\footnote{We use random interleaving in this paper since it is the simplest and most well-known in the literature. Besides, it ensures that the information bits of all devices are assigned random points in M-ary constellation thus guarantying user fairness. Nevertheless, any other interleaving algorithm/array can be used. However, according to used interleaving array, the performance of BIMA may differ. For instance, with the linear interleaving, the performance of one of the devices will be increased whereas for another device it will be degraded. This may cause a slight decay in user fairness performance. The impacts of interleaving algorithms/arrays are left for future works. }. Indeed, a similar unfairness situation can be observed between HP and LP bits in DVB systems once HM is implemented. Therefore, in this paper, we are motivated by data prioritization-based super-constellation of HM in DVB systems. However, we convert it into a novel multiple access scheme and extend to enable massive connectivity in IoT networks. Besides, motivated by the bit-interleaved coded modulation (BICM) \cite{Caire1998} in point-to-point communication, we resolve the unfairness in BIMA. To this end, the proposed BIMA is a novel multiple access scheme which foundations are based on well-matured techniques such as HM and BICM \cite{Caire1998}.  The baseband operations on the transceivers of the proposed BIMA are presented in Fig. 1.
\subsubsection{Power Design} The proposed BIMA has no constraint for the transmit power since no IUI is introduced. There is no restriction either modulations order. According, any number of the IoT devices with any modulation order can be served, unlike conventional NOMA. 

\subsubsection{Signal-to-Noise Ratio (SNR)}
Since only one IQ symbol is transmitted in BIMA, IoT devices are not exposed to additional interference. Thus, only the signal-to-noise ratio (SNR) is defined in BIMA, and it is given for any IoT device as
\begin{equation}
  \mathsf{SNR}_i^{(\mathsf{BIMA})}= \rho|h_i|^2.
\end{equation}
\subsection{Channel Ordering}
In the literature, two different channel ordering schemes have been considered in NOMA systems according to channel state information at the transmitter (CSIT). These are instantaneous channel ordering or statistical channel ordering. In this paper, although channel ordering has no impact on the design of the BIMA, to compare with conventional NOMA schemes, we evaluate the proposed BIMA for both channel ordering schemes. 

\subsubsection{Instantaneous Channel Ordering (ICO)} In this scheme, it is assumed that perfect CSIT is available, and IoT devices are ordered according to their instantaneous CSIT. Hence, considering the PA in conventional NOMA schemes, $|h_1|^2>|h_2|^2,\dots,|h_{L-1}|^2>|h_L|^2$ where  $h_i$ follows $CN(0,\sigma^2), \ \forall i$.
\subsubsection{Statistical Channel Ordering (SCO)}

Since perfect CSIT is not always available, we can order channels according to their second-order statistics (variances), since they change very slowly compared to the instantaneous CSITs and can be obtained with high accuracy. Therefore, in statistical channel ordering (SCO), it is assumed that $\sigma_1^2>\sigma_2^2,\dots,\sigma_{L-1}^2>\sigma_L^2$.

Please note that unlike the common belief for conventional NOMA, according to Shannon's theory, there is no limitation that  $\alpha_i<\alpha_{i+1}$ should be satisfied in case of $|h_1|^2>|h_2|^2,\dots,|h_{L-1}|^2>|h_L|^2$ or $\sigma_1^2>\sigma_2^2,\dots,\sigma_{L-1}^2>\sigma_L^2$. This myth is discussed in detail in \cite{8823873}. However, almost all studies in conventional NOMA assume this order. Therefore, to be in line with the majority of the NOMA studies, we assume the same in this paper.

\section{Performance Analysis}
In conventional NOMA systems, the performance analysis studies are devoted to two categories. The first one is the information-theoretic perspective analysis where the achievable/ergodic rate of a NOMA system is analyzed according to Shannon limit \cite{ShannonC1956} (e.g., by using SINR) and the outage probability of the system is derived. In those studies, none of the base-band operations (e.g., channel coding, IQ modulator/demodulator) is implemented and the analysis is given based on the received SINR at the receiver ends. These studies present the performance limits of the system with infinite inputs (a.k.a., Gaussian inputs). The Shannon's theory shows the maximum achievable performance with ideal techniques but it does not reflect the performance of specific modulation or coding. In other words, the Shannon's formula is a mathematically non-constructive theorem that defines the performance limits (mathematically proven) which  can be achieved by using optimal design but it does not provide what type of signal processing techniques (e.g., coding and modulation) should be used to achieve these limits. For instance, in standards, for data plane, low-density parity-check codes (LPDC) and M-QAM are used for channel coding and modulation/demodulation, respectively. However, it does not mean that the SINR-based performance limits can be achieved with those coding and modulation techniques. On the other hand, in the second category, the bit error rate performance of NOMA systems is analyzed where the NOMA systems take values from finite alphabet inputs (e.g., M-QAM constellation alphabet) with/without a channel coding as being in practical implementations. Therefore, to reflect both perspectives, in this section, we also provide an analysis for BIMA with three key performance indicators (KPIs) (i.e., ergodic capacity, outage probability, and bit error probability) according to both instantaneous and statistical channel ordering.  
\subsection{Ergodic Capacity}

To obtain the ergodic capacity of BIMA, we should firstly derive the individual achievable rate for each IoT device. 
\begin{theorem}
The achievable rate of any IoT device is given by 
\begin{equation}
    R_i=\frac{\log_2\left(M_i\right)}{\log_2\left(M_{bw}\right)}\log_2\left(1+\mathsf{SNR}_i\right), \ i=1,2,\dots,L.
\end{equation}
\end{theorem}
\begin{IEEEproof}
In BIMA, all IoT devices receive one modulated symbol without additional interference. According to the Shannon formula \cite{ShannonC1956}, the achievable rate for a point-to-point communication is given as $\log_2(1+\mathsf{SINR})$. Therefore, the achievable rate of each IoT device can be found as Theorem 2 where the coefficient $\frac{\log_2\left(M_i\right)}{\log_2\left(M_{bw}\right)}$ exists since the $\log_2(M_i)$ bits belong to the $i$th IoT device within the total $\log2(M_{bw})$ bits. 
\end{IEEEproof}
By substituting (13) into (14) and then averaging over channel fading coefficient, the ergodic rate of any IoT device is obtained as
\begin{equation}
        C_i=\frac{\log_2\left(M_i\right)}{\log_2\left(M_{bw}\right)}\int\limits_0^\infty\log_2\left(1+\rho\gamma_i\right)f_{\gamma_i}(\gamma_i)d\gamma_i,
\end{equation}
where $\gamma_i\triangleq|h_i|^2$ and $f_{\gamma_i}()$ is the probability density function (PDF) of $\gamma_i$.

In order to obtain the ergodic rate of the $i$th IoT device, we need to substitute the PDF of $\gamma_i$ into (15), according to which channel ordering scheme is implemented. 
\subsubsection{Instantaneous channel ordering} In case of ICO, the PDF of $\gamma_i$ is equal to the PDF of the $i$th maximum of exponential random i.i.d. $L$ variables. Hence, the PDF of $\gamma_i$ is given by
\begin{equation}
    f_{\gamma_i}(\gamma)=L\binom{L-1}{L-i}\sum_{p=0}^{L-i}(-1)^p\binom{L-i}{p}\frac{1}{\sigma^2}\exp\left(-\frac{\left(i+p\right)\gamma}{\sigma^2}\right).
\end{equation}
\begin{IEEEproof}
See Appendix B.
\end{IEEEproof}

By substituting (16) into (15), and with the help of \cite[eq. (4.337.2)]{Gradshteyn1994}, the ergodic rate of the $i$th IoT device with ICO is derived as
\begin{equation}
\begin{split}
  C_i^{(\mathsf{ICO})}=&-\frac{\log_2(e)\log_2\left(M_i\right)}{\log_2\left(M_{bw}\right)}L\binom{L-1}{L-i} \\
        &\sum_{p=0}^{L-i}(-1)^p\binom{L-i}{p}\frac{1}{i+p}\exp\left(\frac{i+p}{\rho\sigma^2}\right)\mathrm{Ei}\left(-\frac{i+p}{\rho\sigma^2}\right),  
\end{split}
\end{equation}
where $\mathrm{Ei}()$ is the exponential integral function, which is defined as $\mathrm{Ei}(z)=\int\limits_{-z}^\infty\frac{e^{-t}}{t}dt$ \cite[eq. (8.21)]{Gradshteyn1994}.
\subsubsection{Statistical channel ordering} In case of SCO, the PDF of $\gamma_i, \forall i$ has exponential distributions. Hence, by substitution the PDF of exponential distributions (i.e., $f_{\gamma_i}(\gamma)=\frac{1}{\sigma_i^2}\exp\left(-\sfrac{\gamma}{\sigma_i^2}\right)$ ), again with the help of \cite[eq. (4.337.2)]{Gradshteyn1994}, the ergodic rate of the $i$th IoT device with SCO is derived as
\begin{equation}
  C_i^{(\mathsf{SCO})}=-\frac{\log_2(e)\log_2\left(M_i\right)}{\log_2\left(M_{bw}\right)}\exp\left(\frac{1}{\rho\sigma_i^2}\right)\mathrm{Ei}\left(-\frac{1}{\rho\sigma_i^2}\right).
\end{equation}

Finally, the ergodic sum rate of the BIMA for both channel ordering schemes is given as
\begin{equation}
\begin{split}
   & C_{sum}^{(\mathsf{ICO})}=\sum_{i=1}^L C_{i}^{(\mathsf{ICO})}, \\
  &C_{sum}^{(\mathsf{SCO})}=\sum_{i=1}^L C_{i}^{(\mathsf{SCO})}.  
\end{split}
\end{equation}
\subsection{Outage Probability}
The outage event for an IoT device is defined as the probability of the achievable rate being less than the target rate (QoS of that IoT device). This can be defined mathematically as
\begin{equation}
    P_i(out)=P(R_i<\acute{R}_i),
\end{equation}
where $\acute{R}_i$ is the target rate (QoS requirement) of the $i$th IoT device. By substituting (13) and (14) into (20), the OP of the $i$th IoT device is given by
\begin{equation}
    P_i(out)=P(\gamma_i<\phi_i)=F_{\gamma_i}(\phi_i),
\end{equation}
where $\phi=\frac{1}{\rho}\left(2^{\frac{\log_2\left(M_{bw}\right)\acute{R}_i}{\log_2\left(M_{i}\right)}}-1\right)$ is defined and $F_{\gamma_i}(.)$ is the CDF of $\gamma_i$.
\subsubsection{Instantaneous channel ordering} In case of ICO, the CDF of $\gamma_i$ is equal to the CDF of the $i$th maximum of $L$ i.i.d. exponential random variables. Therefore, the OP of the $i$th IoT device is obtained using the CDF of the $i$th maximum of $L$ random variables as 
\begin{equation}
\begin{split}
   &P_i^{(\mathsf{ICO})}(out)= \\
    &\sum_{j=L-i+1}^{L}\sum_{p=0}^{L-j}\binom{L}{j}\binom{L-j}{p}(-1)^p \left(1-\exp(-\frac{\phi_i}{\sigma^2})\right)^{(j+p)}.  
\end{split}
\end{equation}
\begin{IEEEproof}
Please see Appendix C.
\end{IEEEproof}
\subsubsection{Statistical channel ordering} In case of SCO, the CDF of $\gamma_i$ is equal to the CDF of the exponential distribution. Therefore, the probability of outage of the $i$th IoT device is obtained by using the CDF of the exponential distribution (i.e., $F_{\gamma_i}(\gamma)=1-\exp(-\frac{\gamma}{\sigma_i^2})$) as 
\begin{equation}
   P_i^{(\mathsf{SCO})}(out)= 1-\exp(-\frac{\phi_i}{\sigma_i^2}).
  \end{equation}
\subsection{Bit Error Probability Analysis}

To obtain the average bit error probability, we should firstly derive the conditional bit error probability and then obtain average according to channel ordering.
\begin{theorem}
The conditional bit error probability for each IoT device is given as
\begin{equation}
    P_i(e|_{\gamma_i})= \begin{cases} \frac{4\left(\sqrt{M_{bw}}-1\right)}{\sqrt{M_{bw}}\log_2M_{bw}}\mathrm{Q}\left(\frac{3\rho\gamma_i}{M_{bw}-1}\right), \ &\log_4{M_{bw}} \ \text{is integer}, \\ \\
    \frac{4}{\log_2M_{bw}}\mathrm{Q}\left(\frac{3\rho\gamma_i}{M_{bw}-1}\right), \ &\text{otherwise}. \\
    \end{cases}
\end{equation}
\end{theorem}
\begin{IEEEproof}
In the BIMA, thanks to MI at the transmitter, a single IQ symbol is broadcast  to all users as given in (10). Then, each user implements an ML detection (11) and then applies a multiaccess deinterleaving to obtain their own bits. The multiaccess interleaving and deinterleaving are base-band operations in bitwise so they do not affect the error probability of the detection. The conditional error probability is driven by the ML decision. The transmitted symbol to all IoT devices (i.e., $x_{bw}$) turns out to be an $M_{bw}$-ary modulated symbol. Although there is no constraint for the modulation type in BIMA and the system model is presented for arbitrary modulation type in Section II, for BER analysis, we assume that M-QAM is used since it is the most preferred one in the standards. Hence, the error probability of ML decision for a M-QAM signal over a fading channel is given as in Theorem 3 by \cite{Simon2004}. In Theorem 3, the first case defines the square M-QAM when $M_{bw}$ is power of $4$ whereas the second case is given for a rectangular M-QAM. 
\end{IEEEproof}

The average bit error probability is obtained by averaging the conditional bit error probability over $\gamma_i$, which is given by
\begin{equation}
    P_i(e)=\int\limits_0^\infty P_i(e|_{\gamma_i})f_{\gamma_i}(\gamma_i)d\gamma_i.
\end{equation}
By using the alternative representation of $\mathrm{Q}(.)$ function (i.e., $\mathrm{Q}(x)=\frac{1}{\pi}\int_0^{\sfrac{\pi}{2}}\exp\left(-\frac{x^2}{2\sin^2\theta}\right)d\theta$) \cite{Craig1991}, the average bit error probability is given by
\begin{equation}
\begin{split}
   & P_i(e)= \\ &\begin{cases} \frac{4\left(\sqrt{M_{bw}}-1\right)}{\sqrt{M_{bw}}\log_2M_{bw}}\frac{1}{\pi}\int\limits_0^{\sfrac{\pi}{2}}\mathcal{M}_{\gamma_i}\left(-\frac{g}{\sin^2\theta}\right)d\theta,\ \log_4{M_{bw}} \ \text{is integer}, \\ \\
    \frac{4}{\log_2M_{bw}}\frac{1}{\pi}\int\limits_0^{\sfrac{\pi}{2}}\mathcal{M}_{\gamma_i}\left(-\frac{g}{\sin^2\theta}\right)d\theta ,\  \text{otherwise}, \\
    \end{cases}  
\end{split}
\end{equation}
where $g\triangleq\frac{3\rho}{2\left(M_{bw}-1\right)}$. $\mathcal{M}_{\gamma_i}(.)$ is the moment-generating function (MGF) of the random variable $\gamma_i$. 
\subsubsection{Instantaneous channel ordering} In case of ICO, the MGF of $\gamma_i$ is equal to the MGF of the $i$th maximum of $L$ i.i.d. exponentially distributed random variables. Therefore, the average bit error probability of the $i$th IoT device is obtained by using the MGF of the $i$th maximum of L random variables as 
\begin{equation}
\begin{split}
   & P_i^{(\mathsf{ICO})}(e)= \\ &\begin{cases} \Xi_1\frac{L\binom{L-1}{L-i}}{\pi}\int\limits_0^{\sfrac{\pi}{2}}\sum\limits_{p=0}^{L-i}(-1)^p\binom{L-i}{p}\left(\frac{1}{p+i-\frac{g}{\sin^2\theta}\sigma^2}\right)d\theta, \\ \hfill \log_4{M_{bw}} \ \text{is integer}, \\ \\
   \Xi_2\frac{L\binom{L-1}{L-i}}{\pi}\int\limits_0^{\sfrac{\pi}{2}}\sum\limits_{p=0}^{L-i}(-1)^p\binom{L-i}{p}\left(\frac{1}{p+i-\frac{g}{\sin^2\theta}\sigma^2}\right)d\theta  ,\\ \hfill  \text{otherwise}, \\
    \end{cases}  
\end{split}
\end{equation}
where $\Xi_1\triangleq\frac{4\left(\sqrt{M_{bw}}-1\right)}{\sqrt{M_{bw}}\log_2M_{bw}}$  and $\Xi_2\triangleq\frac{4}{\log_2M_{bw}}$ are defined for simplicity of notation. 
\begin{IEEEproof}
See Appendix D.
\end{IEEEproof}

By computing the integral in (27), the average bit error probability is derived in closed form as
\begin{equation}
\begin{split}
    &P_i^{(\mathsf{ICO})}(e)= \\ &\begin{cases} \frac{\Xi_1L}{2}\binom{L-1}{L-i}\sum\limits_{p=0}^{L-i}(-1)^p\binom{L-i}{p}\frac{1}{p+i}\left(1-\sqrt{\frac{g\sigma^2/(p+i)}{1+g\sigma^2/(p+i)}}\right),\\ \hfill \log_4{M_{bw}} \ \text{is integer}, \\ \\
   \frac{\Xi_2L}{2}\binom{L-1}{L-i}\sum\limits_{p=0}^{L-i}(-1)^p\binom{L-i}{p}\frac{1}{p+i}\left(1-\sqrt{\frac{g\sigma^2/(p+i)}{1+g\sigma^2/(p+i)}}\right)  ,\\ \hfill \text{otherwise}.& \\
    \end{cases}  
\end{split}
\end{equation}
\subsubsection{Statistical channel ordering} In case of SCO, the MGF of $\gamma_i$ is equal to the MGF of the exponentially distributed random variable, which is given by
\begin{equation}
    \mathcal{M}_{\gamma_i}(s)=(1+s)^{-1}.
\end{equation}
Hence, substituting (29) into (26), the average bit error probability in case of SCO is derived as
\begin{equation}
    P_i^{(\mathsf{SCO})}(e)= \begin{cases} \frac{\Xi_1}{2}\left(1-\sqrt{\frac{g\sigma_i^2}{1+g\sigma_i^2}}\right),\ \log_4{M_{bw}} \ \text{is integer}, \\ \\
   \frac{\Xi_2}{2}\left(1-\sqrt{\frac{g\sigma_i^2}{1+g\sigma_i^2}}\right) ,\  \text{otherwise}. \\
    \end{cases}  
\end{equation}

\subsection{Benchmark (Conventional NOMA) Performance}
The achievable rate in conventional NOMA is given by \cite{Ding2014} as
\begin{equation}
    R_i^{(\mathsf{conv})}=\log_2{\left(1+\mathsf{SINR}_i^{(\mathsf{conv})}\right)}.
\end{equation}
The ergodic rate of the $i$th IoT device is obtained as 
\begin{equation}
    C_i^{(\mathsf{conv})}=E\left[ R_i^{(\mathsf{conv})}\right],
\end{equation}
where $\mathrm{E}[.]$ is the expectation operator. For both ICO and SCO, the achievable rate is obtained by averaging over instantaneous channel coefficients by using the PDFs 
of related schemes (eq. (16) and the PDF of the exponential distribution for ICO and SCO, respectively). However, averaging rates to obtain EC has more complicated operations, since the imperfect SIC effect $\delta_j\triangleq|x_j-\hat{x}_j|^2$ in (8) is a discrete random variable. For instance, in case of 4-QAM, $\delta_j$ can take values within $[0,2,4]$ with  priori probability of $\mathrm{PEP}(x_j\rightarrow\hat{x}_j)$. $\mathrm{PEP}(x_j\rightarrow\hat{x}_j)$ defines the pairwise error probability when $x_j$ is transmitted and detected as $\hat{x}_j$ for the $j$th IoT device at IoT device $i$.

The outage probability of the $i$th IoT device is given by \cite{Ding2014} as

\begin{equation}
\begin{split}
    & P_i^{(\mathsf{conv})}(out)=P(R_i^{(\mathsf{conv})}<\acute{R}_i) \\      &=
    P\left(\frac{\rho\alpha_i\gamma_i}{\rho\gamma_i\sum\limits_{j=i+1}^L\alpha_j\delta_j+\rho\gamma_i\sum\limits_{p=1}^{i-1}\alpha_p+1}<2^{\acute{R}_i}-1\right).
\end{split}
\end{equation}
With some simplifications, the OP is derived as
\begin{equation}
    P_i^{(\mathsf{conv})}(out)=
    F_{\gamma_i}\left(\frac{ \eta_i\gamma_i}{\rho\left(\alpha_i-\eta_i\left(\sum\limits_{j=i+1}^L\alpha_j\delta_j+\sum\limits_{p=1}^{i-1}\alpha_p\right)\right)}\right).
\end{equation}
where $\eta_i\triangleq 2^{\acute{R}_i}-1$. The outage probability is derived by using the appropriate CDFs according to the channel ordering scheme (eq. (53) and the CDF of exponential distribution for ICO and SCO, respectively). Nevertheless, we again need to consider $\delta_j$ and its priori probabilities $\mathrm{PEP}(x_j\rightarrow\hat{x}_j)$, as explained above.

Although the bit error probability has not been derived for arbitrary modulation orders and numbers of IoT devices in the literature, existing studies \cite{Kara2018f,Bariah2018,Assaf2019,Lee2019,Kara2019,Kara2019a,Garnier2020,NOMA_BER} stipulate that the bit error probability of conventional NOMA is given in the form of
\begin{equation}
    P_i(e|_{\gamma_i})^{(\mathsf{conv})}=\sum_{\lambda=1}^{K}\nu_\lambda \mathrm{Q}\left(\sqrt{\varsigma_\lambda\rho\gamma_i}\right),
\end{equation}
where $K$, $\varsigma_\lambda$, and $\nu_\lambda$ change according to the number of  IoT devices and modulation order. For example, in a network of two-user with $4$-QAM, $K=5$, $\varsigma_\lambda=\sfrac{1}{2}[-1,1,2,1,-1], \ \forall\lambda$ and $\nu_\lambda=[\left(\sqrt{\alpha_2}+\sqrt{\alpha_1}\right)^2,\ \left(\sqrt{\alpha_2}-\sqrt{\alpha_1}\right)^2,\ \alpha_1,  \ \left(2\sqrt{\alpha_2}+\sqrt{\alpha_1}\right)^2,\ \\ \left(2\sqrt{\alpha_2}-\sqrt{\alpha_1}\right)^2]$ and $K=2$, $\varsigma_\lambda=\sfrac{1}{2}, \ \forall\lambda$ and $\nu_\lambda=[\left(\sqrt{\alpha_2}+\sqrt{\alpha_1}\right)^2,\ \left(\sqrt{\alpha_2}-\sqrt{\alpha_1}\right)^2]$ are defined for the first and second IoT devices, respectively \cite{Kara2019}. 

Therefore, the average bit error probability in conventional NOMA is given by
\begin{equation}
    P_i(e)^{(\mathsf{conv})}=\sum_{\lambda=1}^{K}\nu_\lambda\frac{1}{\pi}\int\limits_0^{\sfrac{\pi}{2}}\mathcal{M}_{\gamma_i}\left(-\frac{\varsigma_\lambda\rho}{2\sin^2\theta}\right)d\theta.
\end{equation}

For both channel ordering cases (i.e., ICO and SCO), it can be calculated by using the MGFs of the related ordering given above. 
\section{Fairness}

While improving the overall performance of the system, we also need to consider fairness. Otherwise, one of the users may have performance degradation which causes unfairness among users. In other words, in system design, we cannot only consider overall system performance, but we should also focus on the individual performance of each node. In doing so, we need to ensure that no IoT devices suffers performance degradation. Accordingly, we evaluate the fairness of the proposed BIMA and provide two different fairness metrics to evaluate an IoT device's performance against that of other devices and the overall system.

\subsection {Jain's Fairness Index}

Jain's fairness index is the most common metric for assessing the fairness of a communications system, including NOMA schemes \cite{Di2016,Gui2019}. Jain's fairness index is given as
\begin{equation}
\mathsf{JFI}_\mathsf{C}=\frac{\left(\sum_{i=1}^L R_i\right)^2}{L \sum_{i=1}^L R_i^2},
\end{equation}
where $\mathsf{JFI}_\mathsf{C}$ is defined as the value of the fairness index between 0-1 where 0 represents the most unfair situation and 1 the most fair. In (37), we use the sub-index $\mathsf{C}$ to specify that this fairness index is determined according to the achievable rate/capacity metric. We can easily assess the fairness of both the proposed BIMA and conventional NOMA by substituting (14) and (31) into (37), respectively.

\begin{prop}
Although the fairness evaluation in (37) is the most common metric in NOMA schemes, it does not consider the issue of reliability perspective. It only considers achievable rates; it does not focus on how achievable this communication is or how much is guaranteed. Therefore, we transform Jain's fairness index into alternative forms to represent other KPIs. Considering the outage probability (likewise given in \cite{Liu2020}), we propose a novel JFI that reflects reliability as
\begin{equation}
\mathsf{JFI}_{\mathsf{OP}}=\frac{\left(\sum_{i=1}^L \left(1-P_i(out)\right) \acute{R}_i\right)^2}{ \sum_{i=1}^L \acute{R}_i^2}.
\end{equation}
\end{prop}

In Proposition 1,  $\mathsf{JFI}_{\mathsf{OP}}$ is defined as the value of the fairness index in terms of the OP and the sub-index $\mathsf{OP}$ indicates it. $\acute{R}_i$ determines the target rate for the $i$th IoT device, and $P_i(out)$ is the OP for the given target rate. Substituting (22) and (23) into (38), we can find the fairness of BIMA for the ICO and SCO cases, respectively. The fairness evaluation of conventional NOMA can also be determined by substituting (34) into (38).

\begin{prop}
Likewise, we can further modify the fairness evaluation by also considering the BER metric to evaluate correctly detection of achievable rate. Accordingly, we propose another JFI in terms BER as
\begin{equation}
\mathsf{JFI}_{\mathsf{BER}}=\frac{\left(\sum_{i=1}^L \left(1-P_i(e)\right) \log_2M_i\right)^2}{ \sum_{i=1}^L (\log_2M_i)^2}.
\end{equation}
\end{prop}

In Proposition 2, the sub-index $\mathsf{BER}$ indicates that the fairness is measured by considering the BER performance metric. The fairness index value for BIMA can be obtained by substituting (28) and (29) into (39) for the ICO and SCO cases, respectively. In the same way, the fairness for conventional NOMA is determined by substituting (36) into (39).

\subsection {Proportional Fairness Index}
In the fairness evaluations in (37)-(39), we can measure overall system fairness in terms of capacity, OP, and BER performance metrics. However, we cannot evaluate changes in individual IoT devices; thus, we cannot compare the performance of each IoT device. And we cannot yet ascertain whether any IoT devices experience performance degradation in terms of KPIs. To address this, we need to introduce another fairness index to  compare each IoT devices' performances.  To assess the individual performance, proportional fairness index is commonly used in wireless communications. To this end, we propose following fairness evaluations.

\begin{prop}
We can safely say that the most unfair situation occurs between the IoT devices with the worst and best performances. In this regard, we define proportional fairness index as
\begin{equation}
\mathsf{PFI}_\mathsf{C}=1-\frac{\max\left(R_i\right)-\min\left(R_i\right)}{\max\left(R_i\right)+\min\left(R_i\right)}.
\end{equation}
\end{prop}

In Proposition 3, the sub-index $\mathsf{C}$ again defines that fairness is evaluated in terms of achievable rate/capacity. In (40), $\max{(R_i)}$ and $\min{(R_i)}$ indicate the maximum and minimum achievable rates among IoT devices, respectively. We can obtain the proportional fairness index by substituting related values into (40) for both the proposed BIMA and conventional NOMA (i.e., benchmark). 

\begin{corol}
As we can see in Proposition 3, once the IoT devices perform similarly, the fairness index approaches its best value (i.e., $\mathsf{PFI}_\mathsf{C}\rightarrow 1$) whereas it gets its worst value (i.e., $\mathsf{PFI}_\mathsf{C}\rightarrow 0$) when the devices have large performance gaps. In (40), since we compare the best and worst IoT devices, all other comparisons will have better fairness, so it is not necessary to evaluate others. 
\end{corol}

As with Jain's fairness index, we also need to evaluate the proportional fairness index in terms of other KPIs (i.e., OP and BER). Therefore, we propose two new fairness evaluations in the following.

\begin{prop}
To reflect the reliability of the BIMA, the fairness index in (40) can be defined as
\begin{equation}
\mathsf{PFI}_{\mathsf{OP}}=1-\frac{\max\left(1-P_i(out)\right)-\min\left(1-P_i(out)\right)}{\max\left(1-P_i(out)\right)+\min\left(1-P_i(out)\right)}.
\end{equation}
\end{prop}

\begin{prop}
And to consider the correct detection, in terms of BER, it can be further defined
\begin{equation}
\mathsf{PFI}_{\mathsf{BER}}=1-\frac{\max\left(1-P_i(e)\right)-\min\left(1-P_i(e)\right)}{\max\left(1-P_i(e)\right)+\min\left(1-P_i(e)\right)}.
\end{equation}
\end{prop}

In propositions 4 and 5, the sub-indexes $\mathsf{OP}$ and $\mathsf{BER}$ indicate that the fairness indexes are evaluated in terms of OP and BER performances, respectively. Both fairness values can be easily obtained by substituting related values into (41) and (42) for BIMA and conventional NOMA.

\begin{corol}
We should note that both the OP and the BER expressions are monotonically decreasing functions of SNR (besides they have values less than 1), whereas EC is a monotonically increasing function. Therefore, we modified the values of $\max$ and $\min$ to $\max\left(1-P_i(out)\right)$ and $\min\left(1-P_i(out)\right)$ in (41), and as $\max\left(1-P_i(e)\right)$ and $\min\left(1-P_i(e)\right)$ in (42). As we can see, the value of the proportional fairness index approaches 1 when IoT devices perform similarly, while it approaches to 0 when the performance differs too much in both definitions. 
\end{corol}

\section{Numerical Results}
In this section, we present computer simulations for BIMA in terms of all KPIs (i.e., ergodic capacity, outage probability, and bit error probability) to validate the derived theoretical expressions. Then, we present fairness simulations for BIMA. For the sake of comparisons, in all figures, we also give simulation results for conventional NOMA. In this paper, we consider an ultra-dense network (e.g., IoT application) where the number of devices is massive so that it is not possible to have an orthogonal radio resource block (e.g., time and frequency) for each device. Therefore, inevitably one radio resource block should be shared by multiple devices. In this regard, the comparisons are given for BIMA and NOMA where both need a single resource block to serve multiple users. The comparisons with OMA schemes are beyond the scope of this paper since in that case, we should also consider resource allocation (layer 2) algorithms. 

\renewcommand{\arraystretch}{1.1}
\begin{table}
\centering
\caption{PA of the conventional NOMA in the simulations.}
\begin{tabular}{|c|c||c||c|c|} \hline
&&&\multicolumn{2}{c|}{Commonly Used PA ($\alpha_i$)} \\ \hline
$L$&$M$& Proposed PA ($\alpha_i$)& Values &Ref. \\ \hline \hline
3&4&$0.0261, 0.1948, 0.7791$ & $0.1667, 0.3333, 0.5$&\cite{Ding2014} \\ \cline{1-3}
3&16&$0.0012,0.0588,0.9400$ &$0.05, 0.25, 0.7$ &\cite{Aldababsa2020}\\ \cline{1-3}
3&64&$0.0001, 0.0154, 0.9845$& &\\ \hline
4&4&$0.0063, 0.0473, 0.1893, 0.7571$&$0.1, 0.2, 0.3, 0.4$&\cite{Ding2014} \\ \cline{1-3}
4&16&$0.0001,0.0037,0.0586,0.9377$& $0.02, 0.05, 0.18, 0.75$&\cite{Aldababsa2020}  \\ \cline{1-3}
4&64&$0.0001,0.0037,0.0586,0.9377$& & \\ \hline
\end{tabular}
\label{table1}
\end{table}

\subsection{Evaluation of Proposed PA in Conventional NOMA Networks}
We begin by providing comparisons for conventional NOMA schemes with the proposed PA and commonly used ones in the literature. The proposed PA coefficients of conventional NOMA, which satisfy the constraint in (7), are given in Table I. We also present the commonly used PA for conventional NOMA in the literature.  As we can see, the benchmark PA coefficients are independent from the constellation size and depend only on the number of IoT devices, since all previous PA studies are based on SINR definitions and they do not consider the constellation size. To evaluate the performance of the proposed PA for conventional NOMA, in Figs. 2-4, we present performance comparisons in terms of all KPIs between the proposed and commonly used PA schemes for conventional NOMA. In the comparisons, we assume that $\sigma_i=0$ dB, $\forall i$, and SCO is applied\footnote{The performance comparisons can be extended for various channel conditions and ordering cases. However, the proposed PA in (7) is independent from the channel parameters, and it focuses fundamental theory on reliable SIC detection. Thus, the results will be similar for all channel conditions. Due to space limitations, we skip those comparisons.}. As we can see, in only Fig. 2, where we consider $L=3$ and $M=4$, the proposed PA and the first PA scheme in the literature \cite{Ding2014} perform similarly from an information-theoretic perspective (i.e., EC). However, in terms of the other two critical KPIs (i.e., OP and BER), the proposed PA outperforms the PA in \cite{Ding2014}. This can be explained as follows. The PA in \cite{Ding2014} assumes there to be perfect SIC; however, the realistic imperfect SIC definition in (8), the PA in \cite{Ding2014} becomes ineffective and none of the symbols can be detected at the receivers. On the other hand, the proposed PA and the PA in \cite{Aldababsa2020} perform similarly in Figs. 2-3, where $M=4$ for $L=3,4$, although the proposed PA still induces a slight performance increase. However, the effectiveness of the proposed PA is proved in Fig. 4 with the increase of modulation order $M=16$. In Fig 4, we can see that none of the symbols are detectable with the PA in \cite{Aldababsa2020}, whereas with the proposed PA, we still ensure reliable SIC performance at all receiver ends. 
\begin{figure}
		\centering
    \includegraphics[width=.85\columnwidth]{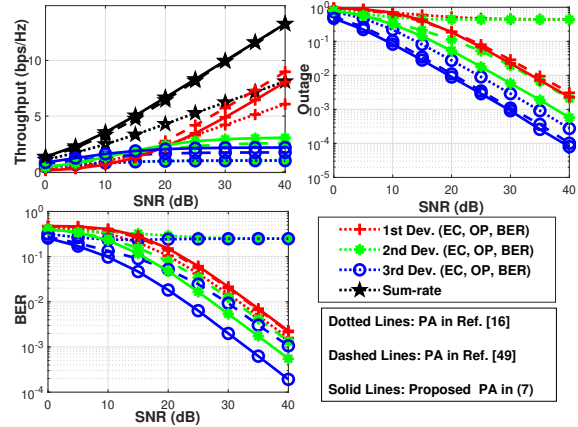}
    \caption{PA comparisons for conventional NOMA when $L=3$, $M_i=4 \ \forall i$ a) EC, b) OP, $\acute{R}_i=\sfrac{M_i}{L}$, c) BER.}
    \label{PA_comp_L_3_M_4}
\end{figure}
\begin{figure}
		\centering
    \includegraphics[width=.85\columnwidth]{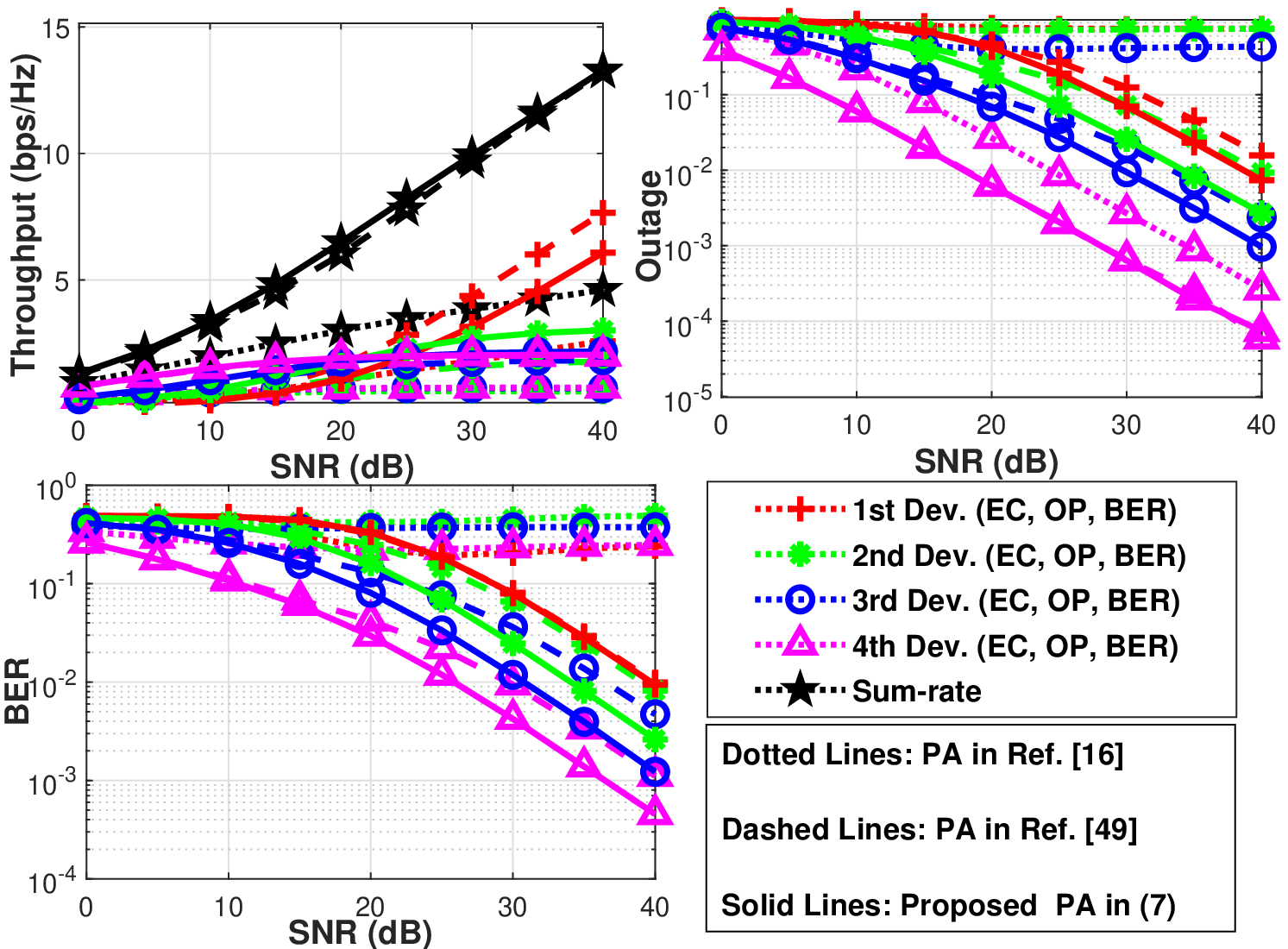}
    \caption{PA comparisons for conventional NOMA when $L=4$, $M_i=4\ \forall i$ a) EC, b) OP, $\acute{R}_i=\sfrac{M_i}{L}$, c) BER.}
    \label{PA_comp_L_4_M_4}
\end{figure}
\begin{figure}
		\centering
    \includegraphics[width=.85\columnwidth]{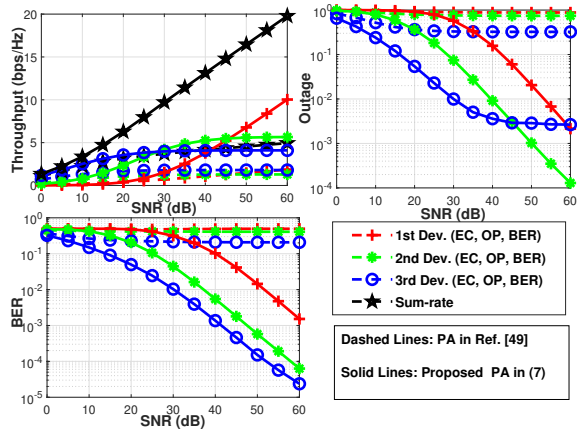}
    \caption{PA comparisons for conventional NOMA when $L=3$, $M_i=16\ \forall i$ a) EC, b) OP, $\acute{R}_i=\sfrac{M_i}{L}$, c) BER.}
    \label{PA_comp_L_3_M_16}
\end{figure}
But, even with the proposed PA, conventional NOMA schemes are limited by the modulation order and the number of IoT devices (which cannot support $M\geq64$ for $L=3,4$ or any modulation order for $L\geq5$). In Table I, we can see that the PA assigned to the first IoT device diminished with increasing numbers of IoT devices or modulation orders. For example, when $L=3$ and $M_i=4 \ \forall i$, the first IoT device gets only $0.0001$ of the total transmit power, which means a performance loss of $40$ dB compared to interference-free systems. This can also be observed for other IoT devices when $L$ is greater (e.g., when $L=3$ and $M_i=4 \ \forall i$ the first and second IoT devices get $\sim1E-6$ and $\sim2E-4$ of the total transmit power, which is equal to performance losses of $60$ dB and $37$ dB, respectively). Furthermore, for $L\geq5$, it is almost impossible to have a proper PA for any modulation order, which limits the usage of conventional NOMA so that it will not allow massive IoT that consists dozens of IoT devices requiring service with the same radio resources. Therefore, the proposed BIMA becomes an enabling technology to support massive IoT networks when more than five IoT devices ($L\geq5$) need support in the same resource block.  As demonstrated in the discussion and figures above, the proposed PA in (7) is the most effective constraint in the literature. Therefore, in the following comparisons, we will use the PA coefficients given in Table I for conventional NOMA performance.

\renewcommand{\arraystretch}{1.1}
\begin{table}
\centering
\caption{Summary of figures in the Numerical Results.}
\begin{tabular}{|c|c||c|c||c|c|} \hline
&& \multicolumn{4}{c|}{Evaluation Metric}\\ \hline
&& \multicolumn{2}{c|}{KPIs (i.e., EC, OP, BER)}& \multicolumn{2}{c|}{Fairness}\\ \hline
$L$&$M$& ICO& SCO&ICO& SCO \\ \hline \hline
3&4&\checkmark (Fig. 5) & \xmark & \xmark& \checkmark (Fig. 11)\\ \hline
3&16& \checkmark (Fig. 6)&\xmark &\xmark& \checkmark (Fig. 12) \\ \hline
4&4& \xmark&\checkmark (Fig. 7) & \checkmark (Fig. 13) & \xmark\\ \hline
4&16& \checkmark (Fig. 8)& \checkmark (Fig. 9) & \checkmark (Fig. 14)&\xmark\\ \hline
5&4& \checkmark (Fig. 10)&\xmark &\xmark& \checkmark (Fig. 15)\\ \hline
\end{tabular}
\label{table2}
\end{table}

\subsection{KPI Comparisons for BIMA and Conventional NOMA}
Comprehensive simulations are presented here for BIMA with various numbers of IoT devices\footnote{Please note that the number of IoT devices refers to the ones occupy/share a single radio resource block. Nevertheless, the BIMA has high scalability and flexibility to support immense IoT networks. For instance, in a large smart factory use-case (e.g., hundreds of IoT devices), BWNOMA can be implemented with a user grouping algorithm where IoT devices in each group (e.g., $L=5$) occupy a single resource block and these groups are separated by orthogonal resource allocation algorithms such as time-division, frequency division or code-division. The user grouping is beyond the scope of this paper and can be investigated in a future publication.} ($ L=3,\ 4,\ 5$) and modulation orders ($ M_i=4,\ 16, \ \forall i$). The simulation results are given for both the ICO and SCO cases. In ICO simulations, we assume that $\sigma_i^2=\sigma^2=0$ dB and the channels are ordered as defined in Section II as $|h_1|^2>|h_2|^2,\dots,|h_{L-1}|^2>|h_L|^2$. On the other hand, in SCO cases, the channels are ordered according to second-order statistics as $\sigma_1^2>\sigma_2^2,\dots,\sigma_{L-1}^2>\sigma_L^2$ where we assume $\sigma_{i+1}^2=\sigma_{i}^2+3$ dB; and $\sigma_L^2=0$ dB.  For the sake of comparison, in all figures, we also present conventional NOMA performance. Please note that as demonstrated in the previous subsection, the proposed PA algorithm outperforms existing PA coefficients in the open literature. Therefore, in the comparisons, we use the proposed PA (i.e., coefficients are given in Table I) for conventional NOMA. It is clear that the performance gain of BIMA would be higher if we had used existing PAs in the open literature such as \cite{Ding2014} or \cite{Aldababsa2020}.  In all the OP comparisons, we set $\acute{R}_i=\frac{M_i}{L}$. In Table II, for ease of follow-up, we summarize the performance comparisons in each figure.\footnote{Due to space limitations, we had to remove the scenarios that are not presented in Table II.} First, regardless of the number of IoT devices, modulation orders, and channel ordering, the derived expressions are perfectly matched with computer simulations of BIMA for all performance metrics. Based on the extensive simulation results, we can deduct the following remarks to evaluate performance of BIMA and to compare those of the conventional NOMA.

\begin{figure}
		\centering
    \includegraphics[width=.85\columnwidth]{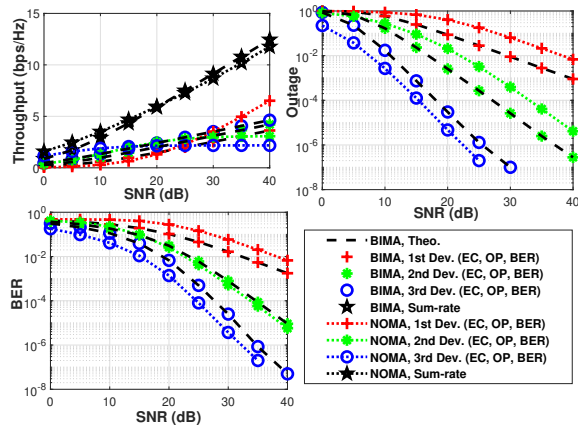}
    \caption{KPIs comparisons for BIMA and conventional NOMA with ICO when $L=3$, $M_i=4\ \forall i$  a) EC, b) OP, $\acute{R}_i=\sfrac{M_i}{L}$, c) BER.}
    \label{KPI_comp_L_3_M_4_ICO}
\end{figure}

\begin{figure}
		\centering
    \includegraphics[width=.85\columnwidth]{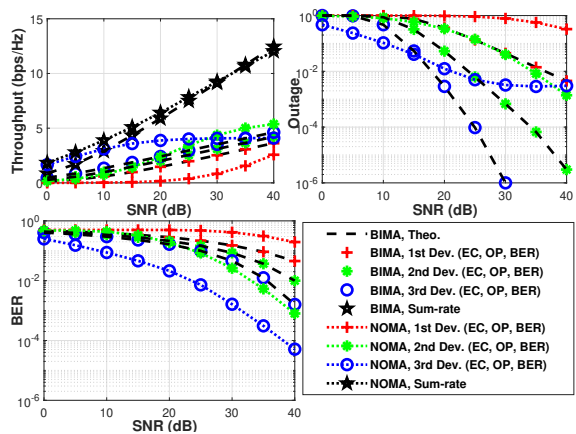}
    \caption{KPIs comparisons for BIMA and conventional NOMA with ICO when $L=3$, $M_i=16\ \forall i$  a) EC, b) OP, $\acute{R}_i=\sfrac{M_i}{L}$, c) BER.}
    \label{KPI_comp_L_3_M_16_ICO}
\end{figure}

\begin{rem}
\textit{BIMA outperforms conventional NOMA in terms of sum-rate (i.e., $C_{sum}\geq C_{sum}^{(\mathsf{conv})}$).  Besides, the achievable rate for each IoT device in BIMA is not limited (i.e., $\lim\limits_{\mathsf{SNR}\rightarrow\infty}C_i\approx\infty$) by interference unlike conventional NOMA where, except for the first IoT device, $\lim\limits_{\mathsf{SNR}\rightarrow\infty}C_i^{\mathsf{(conv)}}=c, \ i\neq1$, where $c$ is a constant which changes according to PA.}
\end{rem}

When comparing the EC curves in all figures, we can add that the proposed BIMA has a capacity performance similar to that of conventional NOMA schemes. Indeed, in terms of sum-rate, in all scenarios, BIMA outperforms conventional NOMA at the mid-high SNR region. This demonstrates the effectiveness of BIMA, which we see that it has a competitive EC performance in addition to its high reliability, low-latency, and improved fairness advantages. In addition, we can observe that the EC for all IoT devices monotonically increases with respect to SNR  in BIMA. However, only the performance of the first IoT device in conventional NOMA has the same behaviour, while the other IoT devices experience a floor that occurs due to IUI. Thus, we can see that even with the best PA selection, except for the first, all other IoT devices in conventional NOMA have limited EC performance even though the transmit power increases. In other words, the increase in the sum-rate of conventional NOMA is only due to the first IoT device, whereas the sum rate in BIMA is shared among IoT devices. This demonstrates the unfairness among IoT devices in conventional NOMA, whereas BIMA provides a fair scheme for all IoT devices. 
\begin{figure}
		\centering
    \includegraphics[width=.85\columnwidth]{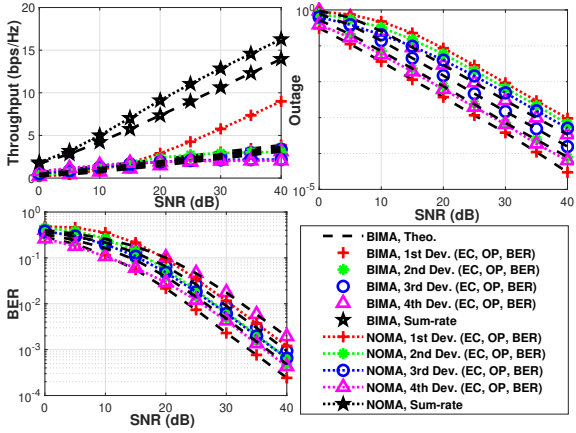}
    \caption{KPIs comparisons for BIMA and conventional NOMA with SCO when $L=4$, $M_i=4\ \forall i$  a) EC, b) OP, $\acute{R}_i=\sfrac{M_i}{L}$, c) BER.}
    \label{KPI_comp_L_4_M_4_SCO}
\end{figure}

\begin{figure}
		\centering
    \includegraphics[width=.85\columnwidth]{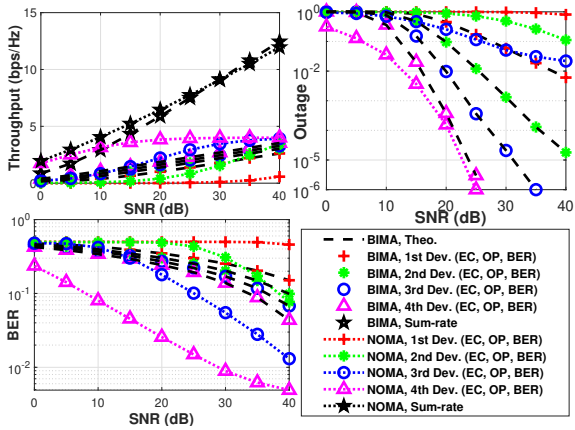}
    \caption{KPIs comparisons for BIMA and conventional NOMA with ICO when $L=4$, $M_i=16\ \forall i$  a) EC, b) OP, $\acute{R}_i=\sfrac{M_i}{L}$, c) BER.}
    \label{KPI_comp_L_4_M_16_ICO}
\end{figure}


\begin{rem}
\textit{BIMA enables massive IoT networks by supporting an arbitrary number of devices and modulation orders. By contrast, conventional NOMA has limitations for both modulation order and number of IoT devices due to strict PA constraint.}
\end{rem}

In all of the figures, we can see that the performance of BIMA increases with respect to SNR for all devices. However, as the number of IoT devices (i.e. $L\geq 3$) and/or modulation order ($M > 4$) increases, some IoT devices are unable to receive service. For example, in Fig. 6, the first IoT device in conventional NOMA is always in outage. This shows that with the increase of QoS requirements (i.e., $\acute{R}$), conventional NOMA can not provide reliable service. In another example, in Fig. 8, not only the first IoT device, but also the second IoT device in conventional NOMA suffers from a performance degradation. Both IoT devices are always out of service (in outage, i.e., $P_i(out)^{(\mathsf{conv})}=1$) until $25$ dB, and then the second IoT device receives service. However, the first IoT device is still in outage. Similarly, both IoT devices also suffer from a lack of BER improvement. The BER performance of second IoT device can only be improved after $25$ dB, whereas for a reliable detection of the first IoT device's symbols, we need $40$ dB SNR. On the other hand, with the proposed BIMA, the same performance metrics can be obtained with $20-30$ dB less transmit power. This is crucial in energy-limited use-cases where it is vital to meet QoS requirements with less power consumption to have a longer life (since in most-cases the devices are powered on batteries and no power grid connection is available). These observations worsen for conventional NOMA with an additional increase in the number of IoT devices or modulation order. However, in BIMA, regardless of the number of IoT devices and modulation order, all IoT devices are guaranteed to have a reliable service. This shows that BIMA ensures QoS requirements. As we can see, in BIMA, none of the IoT devices will experience an outage (i.e. $P_i(out)=1$) or will have a non-detectable symbol (i.e. $P_i(e)=1$).


%

\begin{figure}
		\centering
    \includegraphics[width=.85\columnwidth]{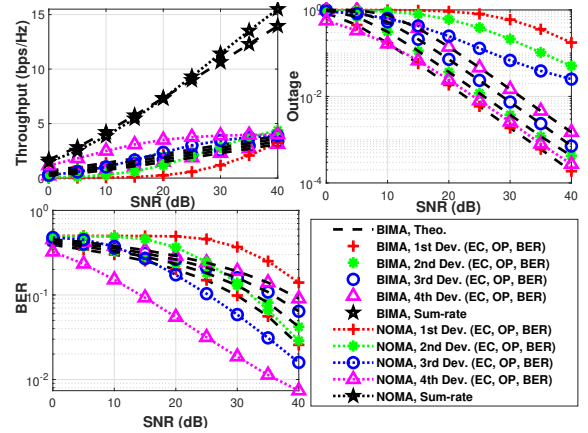}
    \caption{KPIs comparisons for BIMA and conventional NOMA with SCO when $L=4$, $M_i=16\ \forall i$  a) EC, b) OP, $\acute{R}_i=\sfrac{M_i}{L}$, c) BER.}
    \label{KPI_comp_L_4_M_16_SCO}
\end{figure}

\begin{figure}
		\centering
    \includegraphics[width=.85\columnwidth]{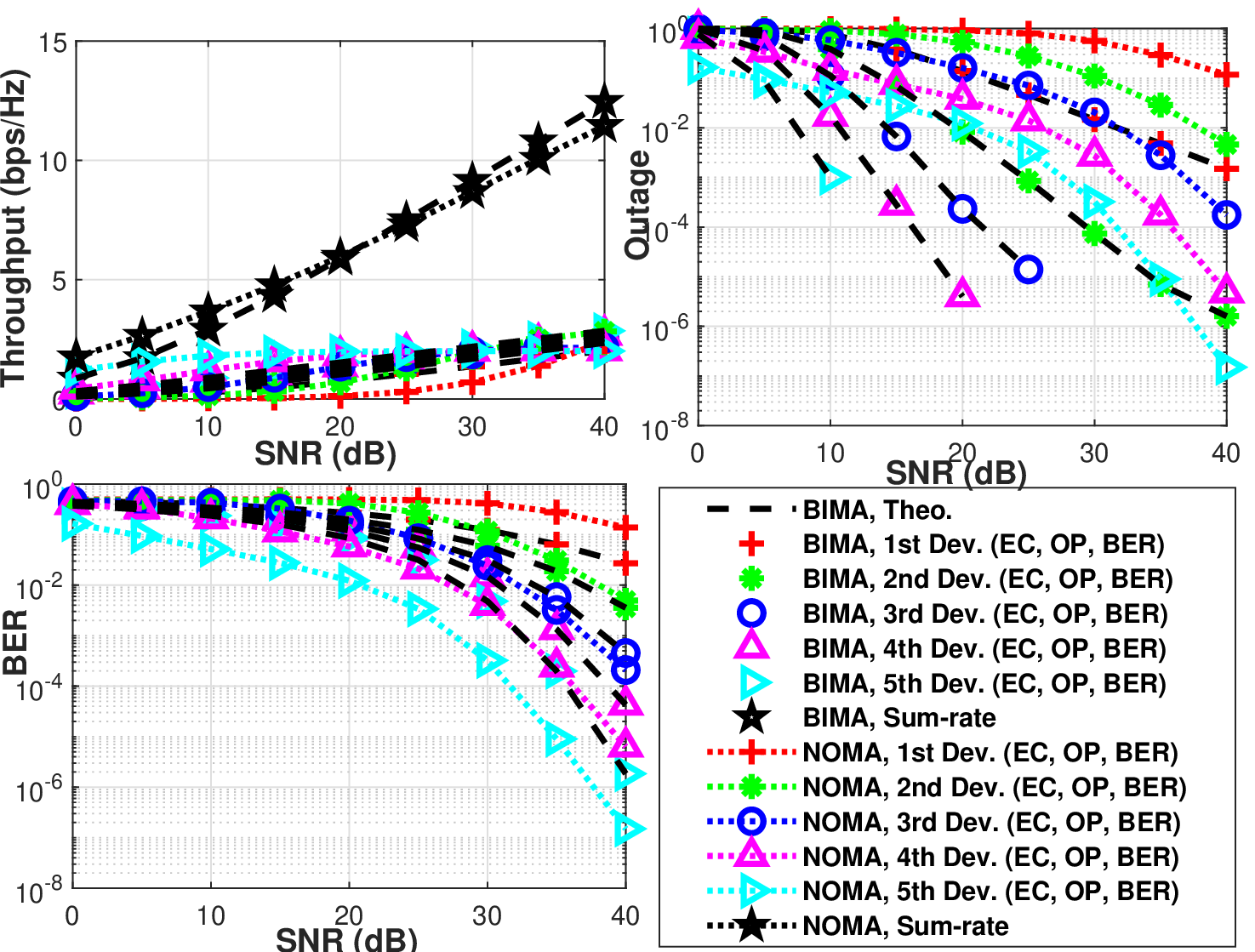}
    \caption{KPIs comparisons for BIMA and conventional NOMA with ICO when $L=5$, $M_i=4\ \forall i$  a) EC, b) OP, $\acute{R}_i=\sfrac{M_i}{L}$, c) BER.}
    \label{KPI_comp_L_5_M_4_ICO}
\end{figure}

\begin{rem}
\textit{ The diversity order for a communications system can be defined in terms of OP  as $\delta=-\lim\limits_{\mathsf{SNR} \rightarrow\infty}\frac{10\log(P_i(out))}{10\log(\mathsf{SNR})}$ or BER  as $\delta=-\lim\limits_{\mathsf{SNR} \rightarrow\infty}\frac{10\log(P_i(e))}{10\log(\mathsf{SNR})}$. The BIMA guarantees the full diversity order. Thus, $ \delta_i^{\mathsf{BIMA}}=L-i+1\ \forall i$ in ICO case and $ \delta_i^{\mathsf{BIMA}}=1\ \forall i$ in SCO case. However, in conventional NOMA, this diversity order can be guaranteed for limited cases. For instance, when $L=4$ and $M=16$, $\delta_i^{\mathsf{(conv)}}\approx0, \ i=1,2$ for both ICO  and SCO cases.} 
\end{rem}

By examining the variable OP and BER performance of BIMA, we can see that as SNR increase, performance also increases. Performance in BIMA never has a floor (e.g., lower floor in error and/or outage). Therefore, BIMA provides the full diversity order for all IoT devices regardless of the scenario. The only change related to diversity order in BIMA is channel ordering. As expected, in the ICO case, since the channels are ordered in an ascending manner, the first IoT device has the diversity order of $L$, while the $L$th IoT device has the diversity order of $1$. However, in conventional NOMA, the QoS requirement cannot always be guaranteed. In particular, some IoT devices may always be out of service or cannot detect their symbols with an increase in SNR. Therefore, there is a floor for some IoT devices in conventional NOMA results (e.g., for the first and third IoT devices in Fig. 6.b, for the first and second IoT devices in Figs. 8.b and 8.c). This error/outage floor means that conventional NOMA has performance limitations, and the performance of some devices cannot be improved even with SNR increase. Thus, the diversity order in those cases is forced to $0$.


\subsection {Fairness Evaluation for BIMA and Conventional NOMA}

As the preceding discussion has shown, BIMA ensures a similar performance and fairness for all IoT devices (e.g., EC is shared equally among IoT devices with a reliable communication). To substantiate this, in Figs. 11-15,  we present the fairness results in terms of both Jain's fairness index and proportional fairness index for various scenarios which are again summarized in Table II. In terms of fairness evaluations, we can deduct the following remarks.

\begin{figure}
		\centering
    \includegraphics[width=0.75\columnwidth]{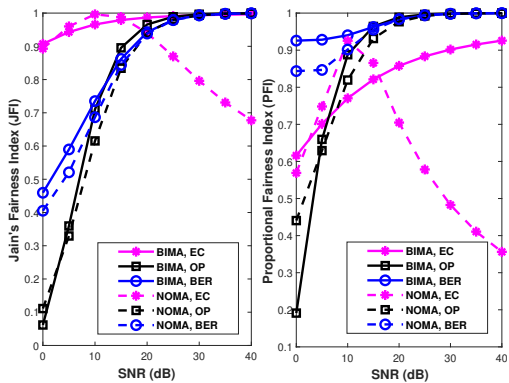}
    \caption{Fairness comparisons for BIMA and conventional NOMA with SCO when $L=3$, $M_i=4\ \forall i$  a) Jain's fairness index (JFI), b) Proportional fairness index (PFI). }
    \label{fairness_comp_L_3_M_4_SCO}
\end{figure}
%

\begin{figure}
		\centering
    \includegraphics[width=0.75\columnwidth]{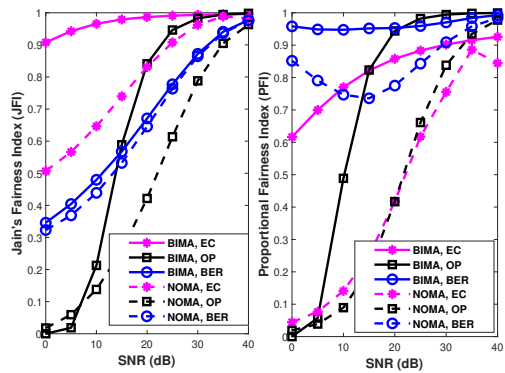}
    \caption{Fairness comparisons for BIMA and conventional NOMA with SCO when $L=3$, $M_i=16\ \forall i$  a) Jain's fairness index (JFI), b) Proportional fairness index (PFI). }
    \label{fairness_comp_L_3_M_16_SCO}
\end{figure}

\begin{rem}
\textit{Regardless of scenarios (e.g., $L$, $M$ or channel ordering), always $\mathsf{JFI}_\mathsf{v}^{\mathsf{BIMA}}>\mathsf{JFI}_\mathsf{v}^{\mathsf{(conv)}}$ and  $\mathsf{PFI}_\mathsf{v}^{\mathsf{BIMA}}>\mathsf{PFI}_\mathsf{v}^{\mathsf{(conv)}}$, where $\mathsf{v}=\mathsf{EC, OP, BER}$.}
\end{rem}

In Figs. 11-15, we can see that in all cases the BIMA offers better fairness than conventional NOMA. Fairness indices for BIMA are invariably higher than those in conventional NOMA. This clearly demonstrates that the BIMA not only provides better KPIs, but also ensures improved user fairness. When we analyze each result in detail, we have the following results. In terms of EC,  the gains of BIMA in Jain's fairness index and proportional fairness index are greater in the case of ICO. This is because in conventional NOMA only the first IoT device has a monotonically increasing performance w.r.t. SNR, whereas all other IoT devices have an upper bound. Furthermore, in ICO, since the first IoT device has the strongest channel gain, the performance gap between IoT devices becomes greater, which causes unfairness. However, in BIMA since all users have a monotonically increasing performance and the total sum-rate is shared in a fair way among the IoT devices, BIMA always offers better fairness. With the increase of $L$ and/or $M$, the fairness gain in BIMA becomes significant compared to conventional NOMA. In Figs. 5-10 we can see that some of the IoT devices in conventional NOMA receive no service (i.e., always experiencing outage) when we have a higher number of IoT devices and/or modulation orders (i.e., higher QoS requirement). Therefore, the reliable bits in the numerators of (38) and (39) decrease, which causes unfairness and a low Jain's fairness index. Likewise, the difference between the minimum and maximum performance in (40)-(42) increases thus causing a worse proportional fairness index. By contrast, in BIMA, all IoT devices perform similarly w.r.t. $L$ and $M$ which provides higher values for both Jain's fairness index and proportional fairness index in terms of all perspectives (i.e., EC, OP, BER).


\begin{figure}
		\centering
    \includegraphics[width=0.75\columnwidth]{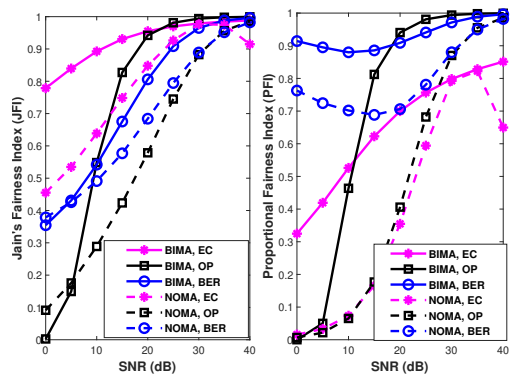}
    \caption{Fairness comparisons for BIMA and conventional NOMA with ICO when $L=4$, $M_i=4\ \forall i$  a) Jain's fairness index (JFI), b) Proportional fairness index (PFI).}
    \label{fairness_comp_L_4_M_4_ICO}
\end{figure}


\begin{figure}
		\centering
    \includegraphics[width=0.75\columnwidth]{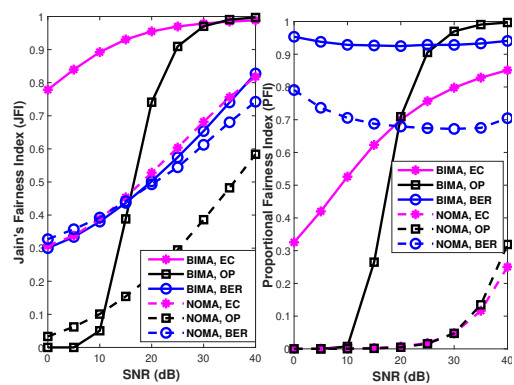}
    \caption{Fairness comparisons for BIMA and conventional NOMA with ICO when $L=4$, $M_i=16\ \forall i$  a) Jain's fairness index (JFI), b) Proportional fairness index (PFI). }
    \label{fairness_comp_L_4_M_16_ICO}
\end{figure}



\begin{rem}
\textit{$\lim\limits_{\mathsf{SNR} \rightarrow\infty} \mathsf{JFI}_\mathsf{v}^{\mathsf{BIMA}}=1$ and $\lim\limits_{\mathsf{SNR} \rightarrow\infty} \mathsf{PFI}_\mathsf{v}^{\mathsf{BIMA}}=1$. However, it cannot be guaranteed in conventional NOMA.}
\end{rem}

As we can see in all of the figures above, as the SNR increases, the fairness indexes for BIMA converge to $1$ (i.e., the best value). BIMA has the best Jain's fairness index and proportional fairness index values in terms of all performance metrics in the high SNR regime. This proves two things: first, that the BIMA provides the overall best fairness performance for all performance metrics, which is the most common evaluation for a communications system; second, that BIMA also guarantees similar performance among IoT devices. Therefore, the IoT devices with the worst and best channel conditions perform similarly, so that none of the IoT devices will have a severe performance. However, in conventional NOMA, we cannot guarantee a convergence in fairness. Indeed, in all figures, the $\mathsf{JFI}_\mathsf{EC}^{\mathsf{(conv)}}$ drops significantly after a critical point and becomes a monotonically  decreasing. The critical point refers to the point where the worst and best EC performance values are the same in Figs.5-10. After that point, the EC performance of the first IoT device in conventional NOMA still increases, whereas the other IoT devices (so the worst EC performance also) encounter an upper bound. Therefore, the proportional fairness index decreases. This can also be observed in the KPI comparisons presented in the previous subsection (Figs. 5-10).
\begin{figure}
		\centering
    \includegraphics[width=0.8\columnwidth]{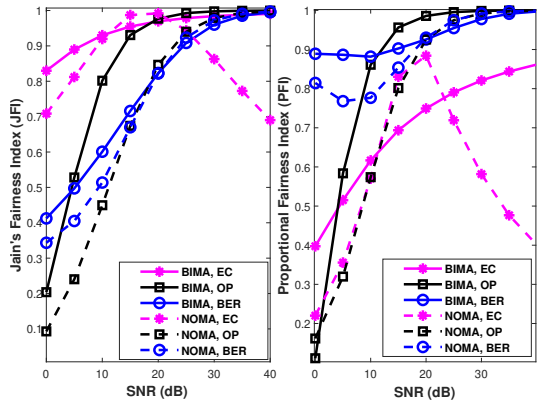}
    \caption{Fairness comparisons for BIMA and conventional NOMA with SCO when $L=5$, $M_i=4\ \forall i$  a) Jain's fairness index (JFI), b) Proportional fairness index (PFI). }
    \label{fairness_comp_L_5_M_4_SCO}
\end{figure}

\section{Complexity and Latency}
At the transmitter side, only mapping operations are required for interleaving and digital baseband modulation blocks. In addition, in conventional NOMA, although multiplications and additions are required for superposition coding (after digital modulation blocks), we neglect the complexity at the transmitter for both the proposed BIMA and conventional NOMA schemes. 

At the receivers, in the proposed BIMA, an ML detector is required for all IoT devices, as explained in Section II. However, in conventional NOMA, the $i$th IoT device should implement $L-i$ times SIC operations before detecting its own symbols. Considering the number of complex operations as the complexity metric, with the aid of \cite{Kim2010}, the complexity of an ML detection is given by 
\begin{equation}
    \mathcal{O}_{\mathsf{ML\ detect}}=4M_{ary},
\end{equation}
where $M_{ary}$ is the modulation order. Hence, the complexity of the BIMA for each IoT device is
\begin{equation}
    \mathcal{O}_{i}^{\mathsf{BIMA}}=4M_{bw},
\end{equation}
and the total complexity of the BIMA is derived as
\begin{equation}
    \mathcal{O}_{\mathsf{BIMA}}=4LM_{bw}.
\end{equation}

By contrast, in conventional NOMA, the SIC should be implemented at the receivers. The SIC includes an ML detector for IoT devices, which are before in the decoding order, and a subtraction from the received signal. The complexity of an SIC operation is given by \cite{Kara2019}
\begin{equation}
    \mathcal{O}_{\mathsf{SIC}}=4M_{ary}+2.
\end{equation}
Therefore, the receiver complexity of the $i$th IoT device in conventional NOMA is given by
\begin{equation}
    \mathcal{O}_{i}^{\mathsf{(conv)}}=4M_i+\sum_{j=i+1}^L(4M_j+2),
\end{equation}
and the total receiver complexity of conventional NOMA is obtained as
\begin{equation}
    \mathcal{O}_{\mathsf{(conv)}}=\sum_{i=1}^{L}4M_i+\sum_{j=i+1}^L(4M_j+2).
\end{equation}
The receiver complexity comparisons between the proposed BIMA and conventional NOMA are given in Table III for different numbers of IoT devices and modulation order. For the sake of comparison, we assume $M_i=M \ \forall i$. In the BIMA, since all IoT devices have the same receiver complexity, the complexity is given for only one IoT device and for the total. On the other hand, in conventional NOMA, results are given for the first IoT device (highest complexity), the last IoT device (lowest) complexity, and the total. As we can see in Table III, with an increase in $L$ and / or $M$, the receiver complexity of conventional NOMA increases linearly, while the complexity of BIMA seems to increase exponentially. This is caused by the total modulation order $M_{bw}$ which increases by the power of 2 as increase in $L$ or $M_i$. However, this increase of the complex operations does not cause additional latency\footnote{This paper focuses on the physical layer aspects of the BIMA scheme. In terms of receiver latency, the proposed BIMA has the same latency as OMA (e.g., TDMA and FDMA) schemes since the detection schemes are quite similar. However, for a fair comparison, the network complexity/latency should be also considered in OMA schemes since (unlike BIMA or conventional NOMA), OMA needs multiple radio resources to serve multiple users. The latency in resource allocation (layer 2) is beyond the scope of this paper and can be considered as future work.} compared to conventional NOMA. The receivers of the BIMA devices are ML detectors; hence, all these operations are computed in parallel and then compared to obtain the minimum.\footnote{Indeed, a single look-up table for an $M_{bw}$-ary constellation can be also used for an ML detection which will have definitely much less latency since only a reading from an address will be needed. Nevertheless, we consider the most complex algorithm for ML detection in comparisons.} However, in conventional NOMA, the IoT devices should compute iterative SIC operations sequentially to detect their own symbols. This causes latency in the receivers. For an informative comparison, in Table IV, we present receiver latency values for the same conditions of $L$ and $M$ given in Table III. The latency values are obtained by implementing ML detectors supported by \textit{MATLAB Communications Toolbox} on the same computer. In the comparisons, the latency values of the BIMA receivers are given in milliseconds, and the values of the conventional NOMA are presented as increases and decreases of percentages compared to the BIMA receiver (negative values mean lower latency, and positive values mean higher latency). For conventional NOMA, we present latency values for the first (highest latency) and last (lowest latency) IoT device, and for the average of all devices. In the BIMA, since all IoT devices have the same latency, the average latency of IoT devices is also equal it. In Table IV, we can clearly see that only the last IoT device in conventional NOMA has lower latency than the IoT devices of the BIMA and this gain is very subtle. By contrast, BIMA provides a lower latency for all other IoT devices and, on average, this latency improvement can achieve a gain of 350\% for the IoT device with the highest latency in conventional NOMA and 170\% for the average latency of IoT devices.  
\begin{table}
\centering
\caption{Receiver complexity of the BIMA and conventional NOMA.}
\begin{tabular}{|c|c||c|c||c|c|c|} \cline{3-7}
\multicolumn{2}{c}{}&\multicolumn{5}{|c|}{Complexity}\\\cline{3-7}
\multicolumn{2}{c|}{}& \multicolumn{2}{|c|}{BIMA}& \multicolumn{3}{c|}{Conv. NOMA}\\ \hline
$L$&$M$&Each IoT dev.&Total&$1$st IoT dev.&$L$th IoT dev.&Total \\\hline \hline
3&2&32&96&28&8&54 \\ \hline
3&4&256&768&52&16&102 \\ \hline
4&2&64&256&38&8&92 \\ \hline
4&4&768&3072&70&16&172 \\ \hline
5&2&128&640&48&8&140 \\ \hline
5&4&4096&20480&88&16&260 \\ \hline
\end{tabular}
\label{table3}
\end{table}

\begin{table}
\centering
\caption{Latency of the BIMA and conventional NOMA.}
\begin{tabular}{|c|c||c||c|c|c|} \cline{3-6}
\multicolumn{2}{c}{}&\multicolumn{4}{|c|}{Latency}\\\cline{3-6}
\multicolumn{2}{c|}{}& BIMA& \multicolumn{3}{c|}{Conv. NOMA}\\ \hline
$L$&$M$&Each IoT dev./Ave.&$1$st IoT dev.&$L$th IoT dev.&Ave. \\\hline \hline
3&2&1.423&$\%172.95$&$-\%9.02$&$\%81.96$ \\ \hline
3&4&1.439&$\%170.33$&$-\%9.89$&$\%80.22$ \\ \hline
4&2&1.435&$\%260.86$&$-\%9.78$&$\%125.54$ \\ \hline
4&4&1.444&$\%259.20$&$-\%10.20$&$\%124.50$ \\ \hline
5&2&1.437&$\%350.45$&$-\%9.91$&$\%170.27$  \\ \hline
5&4&1.451&$\%346.83$&$-\%10.63$&$\%168.10$ \\ \hline
\end{tabular}
\label{table4}
\end{table}

\section{Conclusion}
In this paper, we discussed in detail how to enable massive connections in IoT networks. To this end, we first proposed an efficient power allocation (PA) algorithm for conventional NOMA to guarantee a reliable SIC process, and we demonstrated that conventional NOMA cannot support massive connections, since it has limitations in terms of the number of IoT devices and modulation order. Then, we proposed BIMA to resolve this issue in massive IoT networks. The proposed BIMA has no PA constraint, so it is easy to implement without the need for complex PA optimization algorithms. We provided a comprehensive analytical framework for the proposed BIMA, where EC, OP, and BER analyses were provided for both ICO and SCO cases. We also analyzed the fairness of BIMA in terms of Jain's fairness index and proportional fairness index. Based on extensive computer simulations, we demonstrated that BIMA outperforms conventional NOMA in terms of all KPIs, where BIMA has no limitations in terms of the number of IoT devices and modulation order. This proves that BIMA can enable massive connections in a single resource block, which is crucial for IoT networks. In addition, it was also showed that BIMA provides an improved fairness for the overall system and all IoT devices in terms of all performance metrics. Furthermore, we provided a complexity and latency analysis for BIMA. Detailed analysis and evaluations showed that BIMA supports low-latency communication since it allows parallel computation at the receivers so that no iterative operations are needed (unlike SIC in conventional NOMA). While we introduced BIMA in this paper to enable massive IoT networks, and thus provided analysis for a single antenna case, the applications of BIMA can easily be extended. For instance it can be used for multiple-input multiple-output (MIMO) scenarios since the proposed BIMA is designed to be interference-free so that existing MIMO algorithms can be implemented without further effort. Likewise, the interplay of BIMA with other physical layer techniques can be studied to yield greater performance gains. These represent future directions for further research.

\appendices
\section{Proof of Theorem 1}
\renewcommand{\theequation}{\thesection.\arabic{equation}}
\setcounter{equation}{0}

After the superposition coding at the transmitter, the total constellation is scattered and it does not follow any uniform constellation anymore. Indeed, according to PA coefficients, some points may scatter such a way that these points cannot be recovered even we consider ideal scenario (i.e., no noise, no channel impairments). Therefore, the PA coefficient at the transmitter should be wisely chosen not to cause a non-detectable scattering after superposition coding. To represent this constraints, without loss of generality, in Fig. 16, we present a signal constellation (i.e., the scattered constellation at the transmitter side after superposition coding) of a three-user case with $M_i=4, \ \forall i$ when $b_{3,1}b_{3,2}=00$. At the receiver ends, since $\alpha_3$ has the maximum value within $\alpha_i$, device \#3 implements an ML detection. In ML detection, if the received signal exceeds the decision boundary, detection errors occur. Therefore, none of the symbols, after superposition coding, should cross the decision boundary. Otherwise, even without any impairments (e.g., noise or channel effects), those symbols cannot be recovered. In this case, the decision boundaries for ML detection at the device \#3 for $4$-QAM are the in-phase and quadrature axes. If any of the symbols cross these axes, it means that those symbols are already transmitted erroneously and may not be recovered as original symbols. As given in Fig. 16(a), the constellation points (A, B, C, D, E, F, G) have a higher probability of crossing the decision boundaries (i.e., in-phase or quadrature axes) since they are the closest points to the decision boundaries. Therefore, to detect symbols of device \#3,
\begin{equation}
    \sqrt{\frac{\alpha_3}{2}}>\sqrt{\frac{\alpha_2}{2}}+\sqrt{\frac{\alpha_1}{2}}
\end{equation}
should be satisfied. Otherwise, those constellation points (A,B,C,D,E,F,G) are shifted to other regions of IQ domain at the transmitter side if $\sqrt{\frac{\alpha_3}{2}}\leq\sqrt{\frac{\alpha_2}{2}}+\sqrt{\frac{\alpha_1}{2}}$. In this case, even without any noise or channel effects, these symbols will be detected as $b_{3,1}b_{3,2}\neq00$ which makes certain of erroneous detection. To this end, the PA constraint in (A.1) ensures that none of the points in scattered constellation does not cross any decision boundary of ML decision.
\begin{figure*}[!t]
\centering
\subfloat[Constellation for the 3rd IoT device]{\includegraphics[width=0.85\columnwidth]{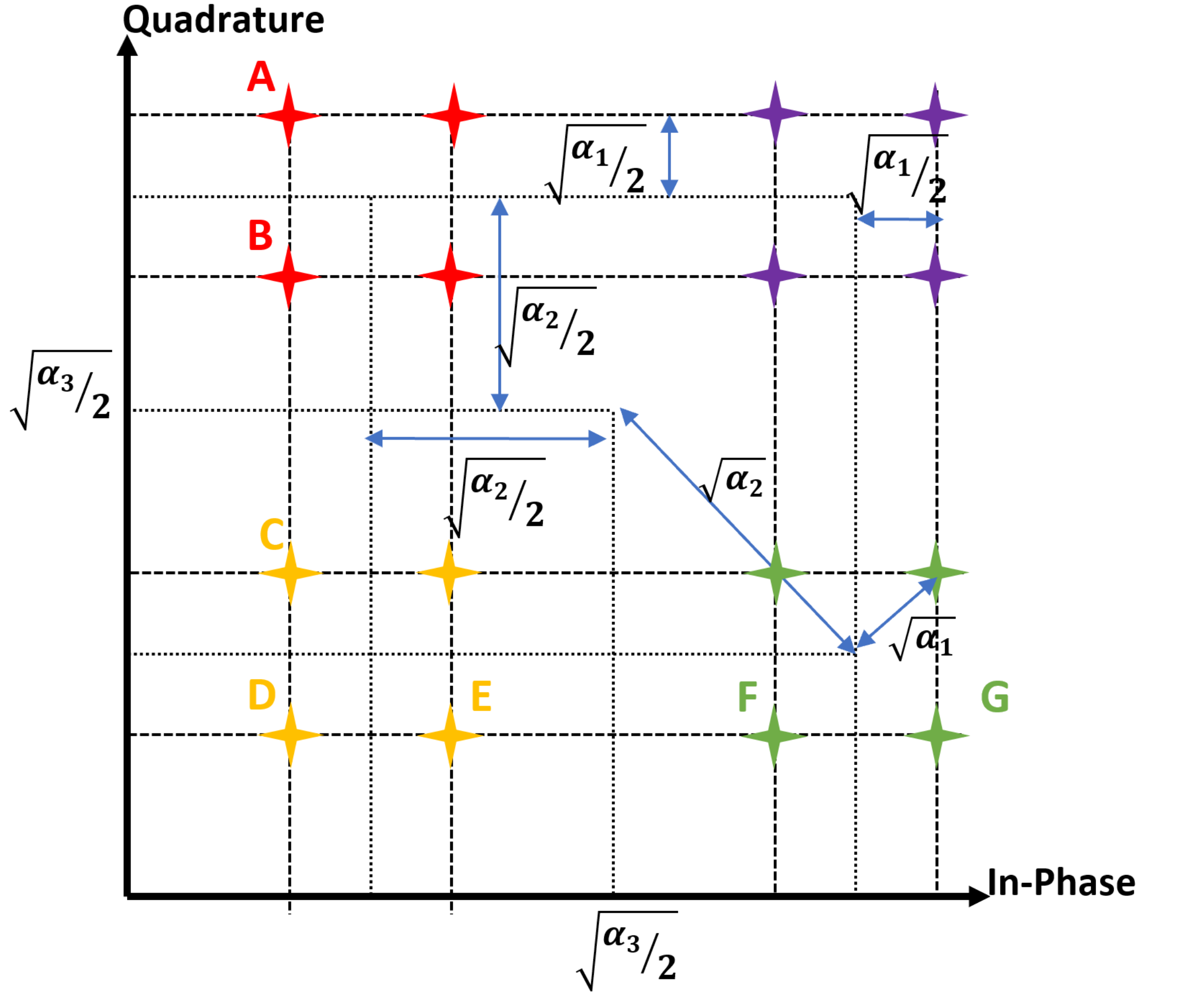}
\label{fig_first_case}}
\hfil
\subfloat[Constellation for the 2nd IoT device after correct SIC]{\includegraphics[width=0.85\columnwidth]{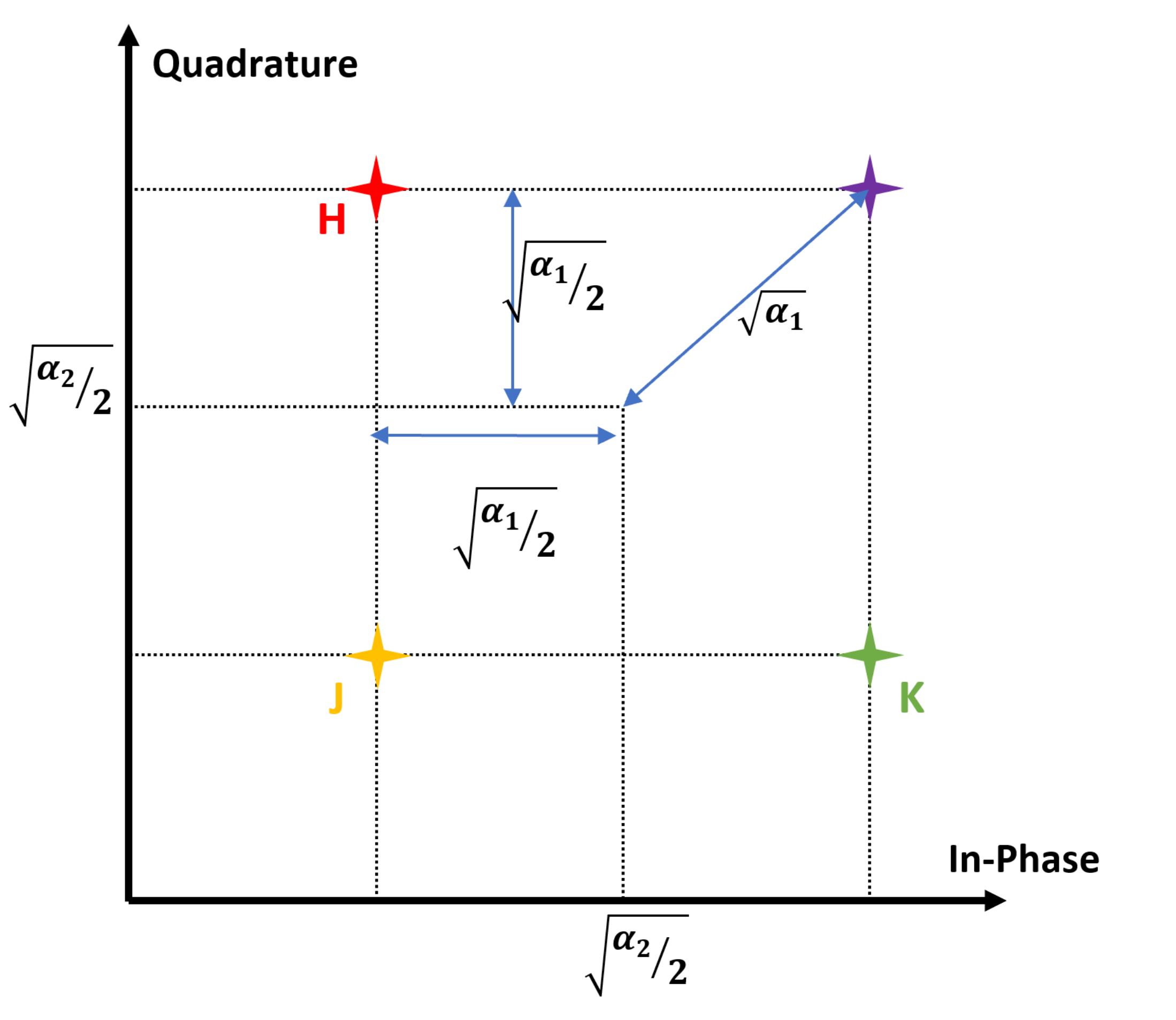}\label{fig_second_Case}}
\caption{Total constellation of conventional NOMA for $L=3$ IoT devices, $M_i=4$-QAM, and $b_{3,1}b_{3,2}=00$.}
\label{fig_sim}
\end{figure*}

Now, let us investigate the PA effects with SIC detection at the receivers. We assume that device \#2 detect the symbols device \#3 correctly (i.e., $b_{3,1}b_{3,2}=00$) and subtracted from the received signal which indicates a correct SIC is implemented at device \#2. In this case, the remaining signal constellation at device \#2 is given in Fig. 16(b). The ML decision boundaries for the symbols of device \#2 are again the in-phase and quadrature axes. Based on the ML decision rule, as discussed above, to detect the symbols of device \# 2 correctly, none of the symbols should cross these boundaries (i.e., other regions of IQ constellation). From Fig. 16(b), we can see that symbols of device \#2 (in H, J, K points) are the most close points to these boundaries. Therefore, we should guarantee that these point not crossing ML decision boundaries and this constraint is given as
\begin{equation}
  \sqrt{\frac{\alpha_2}{2}}>\sqrt{\frac{\alpha_1}{2}}.
\end{equation}
Otherwise, it means that the symbols of device \#2 (in H, J, K points) are already transmitted in a different IQ region and cannot be recovered at the receivers. In this case, regardless of the channel effects and the additive noise, symbols of device cannot be detected. To this end, the constraint in (A.2) should be fulfilled. 

As seen in above discussions with the given constraints, we guarantee at least a detectable scattered constellation at the transmitter side when $L=3$ and $M_i=4, \forall i$. In other words, there will be no conceptual design failure at the transmitter side. At this points, the erroneous detection at the receivers will be affected only the additive noise and channel impairments as being in all communications systems. Now, we extend this rule for general M-QAM constellations with $L$ IoT devices. An $M_i$-QAM modulation can be considered as two $\sqrt{M_i}$-PAM modulations in in-phase and quadrature axes with total $\frac{\mathrm{E}[|x_i|^2]}{2}$ energies. The distance between two adjutant symbols in $\sqrt{M_i}$-PAM modulation is given by
\begin{equation}
    2d_i=2\sqrt{\frac{3}{2\left(M_i-1\right)}}.
\end{equation}
This distance gives how much the adjutant points will be scattered in both in-phase and quadrature axes. Besides, it also determines the distance of a decision boundary from a constellation point (i.e., i.e., $d_i$, half of the distance between two adjutant symbols). 

On the other hand, the maximum distance from the origin of the  $\sqrt{M_i}$-PAM modulation constellation is given by
\begin{equation}
    d_{i,max}=(\sqrt{M}-1)d_i.
\end{equation}
This distance determines the most scattered position for both in-phase and quadrature components after superposition coding. In other words, the points are the most likely crossing the decision boundary (e.g., A, B, C, D, E, F, G, H, J, K points in case of $L=3$, $M_i=4$).

According to these distances, if any of the symbols after superposition coding exceed the decision boundary (i.e., $d_i$, half of the distance between two adjutant symbols), none of the IoT devices will be able to detect its own symbols. Therefore, considering the SIC order, the PA coefficients of $i$th IoT device should be greater than the weighted total of the PA coefficient and $d_{max}$ of the IoT devices that are later in the SIC order. Hence, the PA constraint is given by
\begin{equation}
   \sqrt{\alpha_i}d_i>\sum_{j+1}^{i-1}\sqrt{\alpha_j}d_{j,max}.
\end{equation}
By substituting (A.3) and (A.4) into (A.5), (7a) is obtained. So the proof is completed.


\section{Proof of Eq. (16)}
\setcounter{equation}{0}
We firstly define that $\gamma_i\triangleq\zeta_i$ and assume that $\zeta_1<\zeta_2<\dots<\zeta_n<\dots<\zeta_L$. In this case, $\zeta_n$ is the $n$th minimum of $\zeta_i$ random variables, and according to the order statistics, the PDF of $\zeta_n$ is given by \cite{David2003}
\begin{equation}
    f_{\zeta_n}(\zeta)=Lf_{\zeta}(\zeta)\binom{L-1}{n-1}F_{\zeta}(\zeta)^{(n-1)}\left(1-F_{\zeta}(\zeta)\right)^{(L-n)},
\end{equation}
where $f_{\zeta}()$ and $F_{\zeta}()$ are the PDF and CDF of the i.i.d. $\zeta_i$ random variables. 

With a variable replacement, if we define $i\triangleq L-n+1$, we can easily compute the $i$th maximum of the $\gamma_i$ variables. It is obtained by
\begin{equation}
    f_{\gamma_i}(\gamma)=Lf_{\gamma}(\gamma)\binom{L-1}{L-i}F_{\gamma}(\gamma)^{(L-i)}\left(1-F_{\gamma}(\gamma)\right)^{(i-1)}.
\end{equation}
By substituting PDF and CDF of an exponential distribution, it is derived as
\begin{equation}
    f_{\gamma_i}(\gamma)=L\binom{L-1}{L-i}\frac{1}{\sigma^2}\exp\left(-\sfrac{i\gamma}{\sigma^2}\right)\left(1-\exp\left(-\sfrac{\gamma}{\sigma^2}\right)\right)^{(L-i)}.
\end{equation}
Lastly, by applying the binomial expansion and with some algebraic simplifications, (B.3) is derived as in (16). The proof is completed.
\section{Proof of Eq. (22)}
\setcounter{equation}{0}
As in the previous proof, we define $\gamma_i\triangleq\zeta_i$ and assume that $\zeta_1<\zeta_2<\dots<\zeta_n\dots<\zeta_L$. Hence, the CDF of the $n$th minimum is given by \cite{David2003}
\begin{equation}
   F_{\zeta_n}(\zeta)=\sum_{j=n}^L\binom{L}{j} F_{\zeta}(\zeta)^{(j)}\left(1-F_{\zeta}(\zeta)\right)^{(L-j)}. 
\end{equation}
And again by defining $i\triangleq L-n+1$, the CDF of the $i$th maximum of $\gamma_i$ is obtained by
\begin{equation}
   F_{\gamma_i}(\gamma)=\sum_{j=L-i+1}^L\binom{L}{j} F_{\gamma}(\gamma)^{(j)}\left(1-F_{\gamma}(\gamma)\right)^{(L-j)}.
 \end{equation}
 By applying the binomial expansion and substituting the exponential distribution CDF into (C.2), the CDF of the $i$th IoT device is derived as 
 \begin{equation}
 \begin{split}
    F_{\gamma_i}(\gamma)=
    \sum_{j=L-i+1}^{L}\sum_{p=0}^{L-j}\binom{L}{j}\binom{L-j}{p}(-1)^p &\\ \left(1-\exp(-\frac{\gamma}{\sigma^2})\right)^{(j+p)}.&  
 \end{split}
 \end{equation}
 By suıbstituting the CDF into (21), the outage probability of the $i$th IoT device is derived as (22), so the proof is completed.
 \section{Proof of Eq. (27)}
 \setcounter{equation}{0}
 The moment-generating function of a random variable is given by \cite{Simon2004}
 \begin{equation}
     \mathcal{M}_z(s)=\int_0^\infty f_Z(z)\exp(sz)dz.
 \end{equation}
 By substituting the PDF of the $i$th maximum (16) into (54), the MGF of $\gamma_i$ is given as
\begin{equation}
\begin{split}
   \mathcal{M}_{\gamma_i}(s)=&L\binom{L-1}{L-i}\sum_{p=0}^{L-i}(-1)^p\binom{L-i}{p}\frac{1}{\sigma^2}\\
   &\int\limits_0^\infty\exp\left(-\frac{\left(i+p\right)\gamma}{\sigma^2}\right)\exp(s\gamma)d\gamma.  
\end{split}
\end{equation}
Using the Laplace transform
\begin{equation*}
    \int_0^\infty\exp(-sx)dx=\frac{1}{s},
\end{equation*}
the MGF of $\gamma_i$ is derived as
\begin{equation}
   \mathcal{M}_{\gamma_i}(s)=L\binom{L-1}{L-i}\sum_{p=0}^{L-i}(-1)^p\binom{L-i}{p}\left(p+i+s\sigma^2\right)^{-1}.
\end{equation}
Finally, as defined in (26), we put $s=-\frac{g}{\sin^2\theta}$ into MGF (D.3), the bit error probability is derived as (27), so the proof is completed.


%



\ifCLASSOPTIONcaptionsoff
  \newpage
\fi



%
\bibliographystyle{IEEEtran}
\bibliography{bwnoma_bibtex}
\end{document}